\documentclass[11pt]{article}
\newcommand{\PAPERV}[1]{#1}
\newcommand{\SODAV}[1]{}

\PAPERV{
\usepackage{fullpage,times,amsthm,amssymb,hyperref,graphicx}

\newtheorem{lemma}{Lemma}[section]
\newtheorem{theorem}[lemma]{Theorem}
\newtheorem{corollary}[lemma]{Corollary}
\newtheorem{definition}[lemma]{Definition}
\newcommand{\ssubsection}[1]{\subsection{#1}}
\newcommand{\pparagraph}[1]{\paragraph{#1}}
\newenvironment{pproof}{\begin{proof}}{\end{proof}}

\title{Finding Triangles and Other Small Subgraphs\\
in Geometric Intersection Graphs}
\author{Timothy M. Chan\thanks{Department of Computer Science, University of Illinois at Urbana-Champaign (tmc@illinois.edu).  Work supported in part by NSF Grant CCF-2224271. }}

}

\SODAV{
\usepackage[letterpaper]{geometry}
\usepackage{ltexpprt,amssymb,hyperref,graphicx}

\newtheorem{definition}[lemma]{Definition}
\newcommand{\qed}{\hspace*{\fill}\qquad\vbox{\hrule height0.6pt\hbox{%
   \vrule height1.3ex width0.6pt\hskip0.8ex
   \vrule width0.6pt}\hrule height0.6pt
  }}
\newcommand{\ssubsection}[1]{\subsection{#1.}}
\newcommand{\pparagraph}[1]{\subsubsection*{#1}}
\newenvironment{pproof}{\begin{proof}}{\hspace*{\fill}\end{proof}}

\title{\Large Finding Triangles and Other Small Subgraphs 
in Geometric Intersection Graphs}
\author{Timothy M. Chan\thanks{Department of Computer Science, University of Illinois at Urbana-Champaign (tmc@illinois.edu).  Work supported in part by NSF Grant CCF-2224271. }}

\date{}

\fancyfoot[R]{\scriptsize{Copyright \textcopyright\ 2023 by SIAM\\
Unauthorized reproduction of this article is prohibited}}

}

\newcommand{\IGNORE}[1]{}
\newcommand{\R}{\mathbb{R}}
\newcommand{\eps}{\varepsilon}
\newcommand{\OO}{\widetilde{O}}
\newcommand{\OOO}{O^*}
\newcommand{\D}{\Delta}
\newcommand{\GG}{\widehat{G}}

\begin{document}
\maketitle

\begin{abstract}\SODAV{\small\baselineskip=9pt}
We consider problems related to finding short cycles, small cliques, 
small independent sets, and small subgraphs
in geometric intersection graphs.  We obtain a plethora of new results.
For example:
\begin{itemize}
\item For the intersection graph of $n$ line segments in the plane,
we give algorithms to find a 3-cycle in $O(n^{1.408})$ time,
a size-3 independent set in $O(n^{1.652})$ time,
a 4-clique in near-$O(n^{24/13})$ time,
and a $k$-clique (or any $k$-vertex induced subgraph) in 
$O(n^{0.565k+O(1)})$ time for any constant $k$;
we can also compute the girth in near-$O(n^{3/2})$ time.
\item For the intersection graph of $n$ axis-aligned boxes in
a constant dimension $d$,
we give algorithms to find a 3-cycle in $O(n^{1.408})$ time for any $d$, 
a 4-clique (or any 4-vertex induced subgraph) in 
$O(n^{1.715})$ time for any $d$, 
a size-4 independent set in near-$O(n^{3/2})$ time for any $d$,
a size-5 independent set in near-$O(n^{4/3})$ time for $d=2$,
and a $k$-clique (or any $k$-vertex induced subgraph)
in $O(n^{0.429k+O(1)})$ time for any $d$ and any constant $k$.
\item For the intersection graph of $n$ fat objects in
any constant dimension $d$,
we give an algorithm to find any $k$-vertex (non-induced) subgraph in
$O(n\log n)$ time for any constant $k$, generalizing
a result by Kaplan, Klost, Mulzer, Roddity, Seiferth, and Sharir (1999)
for 3-cycles in 2D disk graphs.
\end{itemize}
A variety of techniques is used, including geometric range searching,
biclique covers, ``high-low'' tricks, graph degeneracy and separators, and shifted quadtrees.
We also prove a near-$\Omega(n^{4/3})$ conditional lower bound for
finding a size-4 independent set for boxes.
\end{abstract}


\section{Introduction}

In computational geometry, many subquadratic-time algorithms have
been developed about finding \emph{pairs} of geometric objects satisfying
some conditions.  For example,
given $n$ red/blue line segments in 2D, 
we can find a bichromatic pair of intersecting segments in
$O(n^{4/3})$ time \cite{Agarwal90a,Matousek93,Chazelle93,ChanZ22}.  Such results are closely related to
geometric range searching \cite{AgaEriSURV,Matousek94survey,BerBOOK,Agarwal17survey}, and research into subquadratic-time algorithms with fractional exponents for different types of geometric ranges has been explored for decades, and has continued to this day (for just one recent example, see~\cite{AgarwalAEKS22}).

But what about problems about finding \emph{triples} of geometric objects
satisfying some pairwise conditions?\footnote{
In other words, we seek a triple $(v_1,v_2,v_3)$ of objects such that some conditions are satisfied concerning the pairs $(v_1,v_2)$, $(v_2,v_3)$, and $(v_1,v_3)$.
In contrast, if the conditions more generally are dependent on the whole triple,
we may encounter 3SUM-hard geometric problems that have near-quadratic
conditional lower bounds (e.g., see~\cite{GajentaanO95,EzraS05}).
}  For example,
given $n$ red/blue/green line segments in the plane, 
how fast can we find a trichromatic triple of line segments
such that every two of them intersect?
In other words, how fast can we find a trichromatic 3-cycle
in the intersection graph?  Here, the \emph{intersection graph} of a
set of geometric objects is the undirected graph in which the vertices are the objects and an edge joins two objects iff they intersect.
(A 3-cycle is also known as a ``triangle'' in graph-theoretic terminology, although
throughout this paper, except in the title, we will use ``3-cycle'' or ``$C_3$'', to avoid confusion with triangles as geometric objects.)
We can ask similar questions about quadruples, e.g., finding 4-cycles ($C_4$) and 4-cliques ($K_4$) in intersection graphs, and $k$-tuples, e.g., finding  $k$-cycles ($C_k$), $k$-cliques ($K_k$), and (induced or not-necessarily-induced) copies of other $k$-vertex subgraphs.

Problems about finding $C_3$ and other small subgraphs for general
dense or sparse graphs have been extensively studied
in the algorithms literature
\cite{AlonYZ97,YusterZ97,KloksKM00,YusterZ04,EisenbrandG04,WilliamsWWY15,DalirrooyfardVW21}
and are well familiar to the SODA audience.
In the computational geometry literature,
there have been works studying certain standard algorithmic problems
on geometric intersection graphs, e.g., connected components \cite{AgarwalK96,Chan06},
all-pairs shortest paths~\cite{ChanS16,ChanS19},  diameter~\cite{BringmannKKNP22}, cuts~\cite{CabelloM21}, and matchings~\cite{BonnetCM20},
but surprisingly not as much on the equally fundamental problems of
finding $C_3$ and other small subgraphs.  One can easily envision 
applications for studying these problems on graphs that are defined
geometrically. For example, cycles with a small number of turns in a road network
correspond to short cycles in a line-segment intersection graph,
and finding a subset in a point set that roughly resembles a fixed pattern
may be formulated as finding a small subgraph in some geometric
graph.  Counting $C_3$'s or small cliques provides a popular statistic about graphs in general, and would also be useful
for geometric graphs that commonly arise in applications such
as unit disk graphs.

\pparagraph{Previous work.}
The problem of finding $C_3$ in geometry-related graphs has been
explicitly addressed in at least three previous papers that we are aware of:

\begin{itemize}
\item Kaplan et al.~\cite{KaplanKMRSS19} in ESA'19 
described an $O(n\log n)$-time algorithm for finding a
3-cycle in the intersection graph of $n$ disks in 2D\@.
(They also considered a certain weighted
variant of the problem, and related problems such as girth.)
\item
Agarwal, Overmars, and Sharir~\cite{AgarwalOS06} in SODA'04
described an $\OOO(n^{4/3})$-time%
\footnote{
Throughout this paper, the $\OOO$ and $\Omega^*$ notation hide $n^\eps$ factors for an arbitrarily small constant $\eps>0$,
and the $\OO$ notation hides $\log^{O(1)}n$ factors.
}
algorithm for finding
a size-3 (or a size-4) independent set in the intersection graph of $n$ unit disks in 2D\@.
This problem is related, since a size-$k$ independent set, which we will
denote by ``$I_k$'', corresponds to a $k$-clique in the complement
of the unit disk graph (and of course, a 3-clique is the same as a 3-cycle).  Alternatively, in the uncomplemented graph, it can be viewed as an induced copy of the graph consisting of $k$ isolated vertices.
(Agarwal et al.'s original motivation was in selecting $k$ 
``maximally separated'' points among $n$ given points.)
\item
Agarwal and Sharir~\cite{AgarwalS02} in SoCG'01 gave an $\OOO(n^{5/3})$-time 
algorithm for finding a congruent (i.e., translated and rotated) copy 
of a fixed triangle $T_0$ among $n$ given points in 3D\@. 
For example, if $T_0$ is the unit equilateral triangle, then
this problem is the same as finding a $C_3$ in the ``unit distance graph''.
Note, though, that this particular problem is very sensitive to precision issues,
since it is about finding an exact match.
(Agarwal and Sharir's paper for the most part was actually devoted to
a \emph{combinatorial} problem, called the
``Erd\H os--Purdy problem'', of bounding the
maximum number of occurrences of a fixed triangle or simplex in a point set.)
\end{itemize}

In combinatorial geometry, there have been a large body of work (e.g., \cite{FoxP08,FoxP12,MustafaP16}) about
$C_3$-free, $K_k$-free, or $K_{k,k}$-free geometric intersection graphs or string graphs, for example, bounding the maximum number of edges, or chromatic number, or establishing the existence of good separators.  However, the algorithmic question of testing
$C_3$-freeness or $K_{k,k}$-freeness was not directly addressed in those
papers (although some of the techniques there are potentially useful, as we will see).

There have been a few results in computational geometry about finding $k$-cliques for general $k$.  For example, Eppstein and Erickson~\cite{EppsteinE94} gave an algorithm
to find a $K_k$ in the intersection graph of $n$ unit disks in 2D
in $O(n\log n + k^2n\log k)$ time (their motivation was in the related
problem of finding $k$ points with minimum diameter among $n$ given points).
For any constant $k$ and constant $d$, they also obtained an $O(n\log n)$-time algorithm for finding
a $K_k$ in the intersection graph of unit balls in $d$D (the hidden dependence on $k$ is exponential).
Finding a $K_k$ in an intersection graph of $n$ boxes%
\footnote{
Throughout this paper, all boxes, rectangles, hypercubes, and squares are axis-aligned.
}
in $d$D
reduces to computing the maximum depth, which can be solved in
$O(n\log n)$ time for $d=2$ and $\OO(n^{d/2})$ time for $d\ge 3$
by techniques for Klee's measure problem~\cite{Chan13}.
On the other hand,
finding a size-$k$ independent set in an intersection graph of $n$ 
objects in 2D such as rectangles and line segments takes $n^{O(\sqrt{k})}$ time by using
planar-graph separators,
or in an intersection graph of $n$ fat objects in $d$D 
in $n^{O(k^{1-1/d})}$ time by using Smith and Wormald's geometric separators
~\cite{SmithW98}.  Nearly matching lower bounds~\cite{Marx06,MarxS14} of the form
$n^{\Omega(k^{1-1/d-\eps})}$ are known even for 2D unit squares or unit disks, assuming the Exponential Time Hypothesis (ETH), implying that such problems
are not fixed-parameter tractable with respect to $k$.
(There is also an extensive body of work on \emph{approximation} algorithms
for independent set and clique in geometric intersection graphs, which are not directly relevant here and will be ignored.)

Despite all the works noted above, surprisingly there were still no subquadratic algorithms known (to the best of our knowledge) for finding the simplest
patterns, like $C_3$, in intersection graphs of line segments.  This paper will rectify the situation.



\pparagraph{New results.}
We obtain a plethora of new results, as summarized in Table~\ref{tbl}.
The 23 new upper bounds are derived from 19 different algorithms or theorems/corollaries!
We focus only on the detection versions of these problems: deciding whether a pattern occurs, and if so, reporting one occurrence.
We focus on the most basic types of geometric objects, namely,
boxes in $d$D, line segments in 2D, and fat objects
in $d$D; throughout this paper, the dimension $d$ is always considered a constant.
Though not necessarily indicated in the table, some (but not all) of the results hold for more general classes of geometric graphs, and 
for colored variants of the problems; readers are referred to the theorem statements for the details.

\begin{table}[t]
\centering
\begin{tabular}{|l|l|lll|}\hline
objects & pattern& run time & ref. & notes \\\hline
boxes in $d$D & $C_3$ & $O(n^{1.408})$ & Thm.~\ref{thm:boxes:Ck}
& {\small ($\Omega^*(n^{4/3})$\ \ Thm.~\ref{thm:boxes:C3:lb})}$\!\!$ \\
      & $I_3$ & $\OO(n)$ & Thm.~\ref{thm:boxes:I3} &\\
      & $C_4$ & $\OO(n)$ & Thm.~\ref{thm:boxes:Ck:even} &\\
      & $K_4$ or any induced $X_4$ & $O(n^{1.715})$ & Thm.~\ref{thm:boxes:K4} &\\
      & $I_4$ & $\OO(n^{3/2})$ & Thm.~\ref{thm:boxes:I4}
& {\small ($\Omega^*(n^{4/3})$\ \ Thm.~\ref{thm:boxes:I4:lb})}$\!\!$\\
      & $C_k$ for $k\equiv 0\bmod 4$ & $\OO(n^{2-4/k})$ &  Thm.~\ref{thm:boxes:Ck:even} &\\
      & $C_k$ for $k\equiv 2\bmod 4$ & $\OO(n^{2-4/(k+2)})$ &  Thm.~\ref{thm:boxes:Ck:even} &\\
      & $C_k$ for $k$ odd & 
$O(n^{2-2/(k+1)})$ 
& Thm.~\ref{thm:boxes:Ck} &\\
      & $K_k$, $I_k$, or any induced $X_k$ & $O(n^{0.429k+O(1)})$ & Cor.~\ref{cor:boxes:Kk} &\\
boxes in 5D & $I_4$ & $\OO(n)$ & Thm.~\ref{thm:boxes:I4:5d} &\\
boxes in 2D & $I_5$ & $\OO(n^{4/3})$ & Thm.~\ref{thm:boxes:I5:2d} &\\
\hline
line segments in 2D$\!\!$ & $C_3$ & $O(n^{1.408})$ & Thm.~\ref{thm:segs:Ck}
&\\
& $I_3$ & $O(n^{1.652})$ & Thm.~\ref{thm:segs:I3} &\\
& trichromatic $C_3$ & $\OOO(n^{7/4})$ & Thm.~\ref{thm:segs:C3} &\\
& $C_4$ & $\OO(n)$ & Thm.~\ref{thm:segs:C4} &\\
& $K_4$ & $O^*(n^{24/13})$ & Thm.~\ref{thm:segs:K4} &\\
& $C_k$ for $k$ even & $O(n^{1.187})$ & Thm.~\ref{thm:segs:Ck:even:sep} &\\
& $C_k$ for $k$ odd & 
$O(n^{2-2/(k+1)})$ 
& Thm.~\ref{thm:segs:Ck} &\\
& girth & $\OO(n^{3/2})$ & Thm.~\ref{thm:segs:girth}&\\
& $K_k$ or any induced $X_k$ & $O(n^{0.565k+O(1)})$ &
Cor.~\ref{cor:segs:Kk} &\\
\hline
fat objects in $d$D & $C_k$, $K_k$, 
or  any (non-induced) $X_k$$\!\!$ & $O(n\log n)$ & Thm.~\ref{thm:fat} &\\ 
\hline
translates in 2D & $C_3$ or $I_3$ & $\OO(n^{3/2})$ & Thm.~\ref{thm:translates} &\\
translates in 3D & $C_3$ or $I_3$ & $\OO(n^{9/5})$ & Thm.~\ref{thm:translates} &\\
\hline
\end{tabular}
\IGNORE{
K_3: n^{1.408}     I_3: n^{1.652}   #K_3: n^{7/4}
K_4: n^{24/13}
C_4: n
C_k: n^{1.187} if k even, n^{< 2-2/(k+1)} if k odd
girth: n^{3/2}   (or n^{1.402} + kn^{1.187})
K_k or I_k or any k-node induced subgraph: n^{0.55??k+...} 

}
\caption{Time complexity for finding a small subgraph in a geometric intersection graph.  Here, $d$ and $k$ are constants;
$C_k$ denotes a $k$-cycle;
$I_k$ denotes a size-$k$ independent set; $K_k$ denotes a $k$-clique;
and $X_k$ denotes an arbitrary fixed $k$-vertex subgraph.
``Translates'' refer to translated copies of a fixed geometric object
of constant description complexity.
}
\label{tbl}
\end{table}

As can be seen from the table, we have identified a rich class of
problems in computational geometry that are solvable in polynomial time with
interesting exponents, going beyond traditional problems about
finding pairs of geometric objects.
As one can predict, some of these fractional exponents come from
the use of geometric range searching, but in far less obvious ways.

A little more surprising, at least at first glance, is
that unusual exponents arise even in the case of boxes (e.g., see 
our results for $K_4$ and $I_4$),
as problems about boxes typically require only \emph{orthogonal} range searching, which are
easier and admit polylogarithmic-time data structures.
On the other hand, the aforementioned known results about fixed parameter intractability
in terms of $k$, which hold even for 2D unit squares, indicate that 
time bounds for $I_k$ eventually have to be super-linear as the constant $k$ gets sufficiently large (though with Marx et al.'s intractability proofs \cite{Marx06,MarxS14}, it seems $k$
needs to be fairly large).
We prove that finding $C_3$ for 3D boxes
or $I_4$ for 6D boxes already requires $\Omega^*(n^{4/3})$ time,
under certain hypotheses from fine-grained complexity.
This lower bound for $I_4$ in 6D complements nicely with the
near-linear upper bounds for $I_3$ for any dimension $d$, as well as
for $I_4$ in 5D.

Some of our algorithms use fast
(square or rectangular) matrix multiplication, but some do not.
Even for those that use it, weaker but still nontrivial upper bounds
can be obtained with naive matrix multiplication.
See some of the theorem statements for the time
bounds expressed in terms of the matrix multiplication exponent $\omega<2.373$~\cite{AlmanW21}.
There have been several previous examples of the usage of fast matrix multiplication in computational geometry,
e.g., on dynamic geometric connectivity \cite{Chan06,ChanPR11}, 
shortest paths in geometrically weighted graphs \cite{Chan10apsp},
colored orthogonal range counting \cite{KaplanRSV08}, \ldots, and our work
may be viewed as a continuation of this interesting line of research.

Our results on $K_4$ for boxes and line segments should be compared with 
the current best time bound on $K_4$ for general graphs, which
is $O(n^{\omega(1,2,1)})\le O(n^{3.252})$~\cite{EisenbrandG04}.\footnote{
Here, $\omega(a,b,c)$ denotes the rectangular matrix multiplication exponent.
In other words, let $M(n_1,n_2,n_3)$ be the complexity of multiplying an $n_1\times n_2$ with an $n_2\times n_3$ matrix;
we have $M(n^a,n^b,n^c)=\Theta^*(n^{\omega(a,b,c)})$.
}
Our results on $K_k$ for boxes and line segments should be compared
with the current best bound on $K_k$ for general graphs,
which is about $O(n^{k\omega/3+O(1)})\le O(n^{0.791k+O(1)})$~\cite{EisenbrandG04}.  Our algorithm on $K_k$ for boxes is
faster than the aforementioned $\OO(n^{d/2})$ algorithms if
$k$ is small relative to $d$.  While $n^{O(\sqrt{k})}$ algorithms
were known for $I_k$ for 2D boxes and line segments as mentioned,
no $n^{o(k)}$ algorithms were known for $I_k$, e.g., for arbitrary 3D boxes.

Our $\OO(n^{3/2})$ upper bound for computing the \emph{girth} (i.e., the shortest cycle length) for line-segment intersection graphs contrasts with
a recent near-quadratic lower bound on computing the \emph{diameter}
for line-segment intersection graphs by Bringmann et al.~\cite{BringmannKKNP22}.

Our $O(n\log n)$ algorithm for finding a (not-necessarily induced) copy of any
fixed $k$-vertex subgraph in intersection graphs of fat objects in $d$D,
for any constant $k$ and $d$, greatly generalizes 
the recent result by Kaplan et al.~\cite{KaplanKMRSS19}, which was only for
finding $C_3$ in the intersection graph of disks in $\R^2$.
It is also interesting to compare our result with Eppstein's well-known
result on finding a fixed subgraph in planar graphs in linear time~\cite{Eppstein99} from SODA'95: on one hand, every
planar graph can be realized as an
intersection graphs of 2D disks  by the Koebe--Andreev--Thurston
theorem; on the other hand, all of our algorithms require a geometric
realization to be given.

\pparagraph{Techniques.}
For our class of problems, there is actually an abundance of available
techniques, and many different ways to combine them.  
Our contributions lie in collecting all of these techniques together,
and in finding the best combination to achieve each of the individual results.
We will organize the paper by techniques, 
which will hopefully help researchers in the future find further
improvements and solve more problems of this kind.

One immediate challenge is that intersection graphs may be
dense, and so to obtain subquadratic results
e.g.\ for $C_3$, we can't afford to construct the intersection graph explicitly.
In Section~\ref{sec:bicliques}, we describe a known
technique, \emph{biclique covers}
(closely related to range-searching data structures),
which provides one convenient way to compactly represent intersection graphs.
Biclique covers have been used before to directly reduce problems
about geometric intersection graphs to problems on non-geometric
sparse graphs; for example, see \cite{Chan06} on geometric connectivity.
Some readers may have recognized that the $O(n^{1.408})$ bounds
in Table~\ref{tbl} for $C_3$ are the same as the current known bounds
for $C_3$ in sparse graphs~\cite{AlonYZ97}.  Indeed, with biclique covers, we 
give an easy black-box reduction of $C_k$ for boxes (or monochromatic line segments) to $C_k$ in sparse graphs.
However, the reduction works only for $C_k$, but not for $K_k$ or
other subgraphs.  Also, the reduction doesn't work as well for many
types of non-orthogonal objects, such as trichromatic line segments, because the biclique
cover complexity is not small enough.  Thus, one can't just blindly
apply known results for general sparse graphs to solve geometric problems  involving other patterns and other types of objects.

In Section~\ref{sec:high:low}, we combine biclique covers and range searching with another technique that may be dubbed ``high-low tricks''.  
Many of the known algorithms for finding small subgraphs in sparse graphs
work by dividing into cases based on whether vertices
have high degree or low degree.  We will do something similar, but
instead classifying \emph{edges} as high or low.
We describe a multitude of new algorithms based on this idea.
This is the most technically sophisticated part of the paper,
as there are multiple possiblilities on how to divide into cases and how to
address each case, especially for more complicated patterns such as $K_4$.

In Section~\ref{sec:deg}, we exploit the observation that for
certain types of geometric objects and certain patterns such as $C_4$, the pattern must
always occur unless the intersection graph is sparse.  But if graph is 
sparse and, more precisely, has low \emph{degeneracy}, faster algorithms exist.
In Section~\ref{sec:sep}, we further use the fact that for the
line segment case in the plane, if the intersection graph is sparse, it must have small
\emph{separators}---this enables efficient divide-and-conquer algorithms.
Results in Sections~\ref{sec:deg} and~\ref{sec:sep} are less general
(for example, they do not apply to colored variants of the problems),
although some might actually prefer algorithms that exploit fully the
geometry of
the specific problems.

In Section~\ref{sec:quadtree}, we present a different approach for fat
objects, by using \emph{shifted quadtrees} plus dynamic programming.  This is more general
than Kaplan et al.'s previous approach for disks~\cite{KaplanKMRSS19},
and conceptually simpler than Eppstein's for planar graphs~\cite{Eppstein99} as well.  

Lastly, in Section~\ref{sec:recurs}, we present yet another approach: a \emph{round-robin recursion} using geometric cuttings,
based on one of
Agarwal and Sharir's proofs on the Erd\H os--Purdy problem~\cite{AgarwalS02}.  
We show that their technique, originally devised for a combinatorial geometry problem,
can also be used to obtain new algorithms.
However, this technique seems more limited, and currently is 
responsible only for the results on 2D and 3D translates
in Table~\ref{tbl}.

We begin with conditional lower bounds in Section~\ref{sec:lb}, as the proofs
are short. In particular, the lower bound proof for $I_4$ is neat; it is based on the \emph{Hyperclique Hypothesis}~\cite{LincolnWW18},
which has started to gain prominence in fine-grained complexity
and computational geometry (e.g., see~\cite{BringmannKKNP22}).


\section{Conditional Lower Bounds}\label{sec:lb}

For certain versions of our problems involving nonorthogonal objects,
e.g., trichromatic $C_3$ for line segments, it is not difficult
to obtain super-linear conditional lower bounds.
For example, the well-known \emph{Hopcroft's problem}
(deciding whether there is a point-line incidence among $n$ points
and $n$ lines in 2D) is generally believed to require
$\Omega^*(n^{4/3})$ time~\cite{Erickson95,Erickson96}, and it is straightforward to reduce Hopcroft's problem
to trichromatic $C_3$ for line segments.  In this section, we obtain
super-linear conditional lower bounds even for versions of the problems
for \emph{orthogonal} objects, namely, boxes, and even for very small $k$.

\ssubsection{$C_3$ in 3D box intersection graphs}

To warm up, we note an easy reduction from $C_3$ in sparse graphs to 
$C_3$ in box intersection graphs.  It has been conjectured that
finding a $C_3$ in a graph with $m$ edges requires $\Omega^*(m^{4/3})$ time~\cite{AbboudW14,AbboudBKZ22},
and so the reduction implies an $\Omega^*(n^{4/3})$ lower bound for
$C_3$ in box intersection graphs under this conjecture.
(The running time of the fastest known algorithm~\cite{AlonYZ97}
is $O(m^{2\omega/(\omega+1)})\le O(m^{1.408})$ under the current matrix multiplication exponent bound $\omega<2.373$, and is $\OOO(m^{4/3})$ if $\omega=2$.
A stronger ``all-edges'' variant of the $C_3$ finding problem is known to have
an $\Omega^*(m^{4/3})$ lower bound under more standard conjectures,
such as the 3SUM Hypothesis and the APSP Hypothesis~\cite{Patrascu10,WilliamsX20,ChanWX22}.)
The reduction below is similar to a known
conditional lower bound proof for Klee's measure problem in the
3D case~\cite{Chan10}.

\begin{theorem}\label{thm:boxes:C3:lb}
If there is an algorithm for finding a $3$-cycle
of $n$ boxes in $\R^3$ in $T(n)$ time,
then there is an algorithm for finding a $3$-cycle
in a sparse directed graph with $m$ edges in $O(T(m))$ time.
\end{theorem}
\begin{pproof}
We are given a directed graph $G=(V,E)$ with $m$ edges.
Assume the vertices are numbered $\{1,2,\ldots,n\}$.
Create the following boxes in $\R^3$, for each $(u,v)\in E$:
$\{u\}\times \{v\}\times\R$, and
$\R\times \{u\}\times \{v\}$, and
$\{v\}\times\R\times \{u\}$.  (These boxes are technically axis-parallel lines, but can be
thickened to be fully 3-dimensional.)
It is easy to see that
the intersection graph of these $O(m)$ boxes has a 3-cycle iff $G$ has a 3-cycle.
\end{pproof}

Later in Theorem~\ref{thm:boxes:Ck}, we will prove a reduction in the opposite direction, thus showing equivalence (up to polylogarithmic factors).

\ssubsection{$I_4$ in 6D box intersection graphs}

For independent set, we prove an $\Omega^*(n^{4/3})$ lower bound 
for size 4, under a different conjecture, that
finding a size-$k'$ hyperclique in a $k''$-uniform hypergraph with $N$ vertices
requires $\Omega^*(N^{k'})$ time, for any constants $k'>k''\ge 3$.
This conjecture is known as the \emph{Hyperclique Hypothesis}~\cite{LincolnWW18}
and has been used in a few recent papers (e.g., \cite{BringmannKKNP22}).
In our reduction, we only need the case $k'=4$ and $k''=3$.
Our reduction works for orthants, i.e., $d$-dimensional boxes with $d$ unbounded
sides, one per coordinate axis. (Thus, our reduction also works for unit hypercubes, since
we can replace orthants with sufficiently large congruent hypercubes, and then rescale).

As we will see, part of the challenge in devising the reduction is in enforcing equalities of certain indices.  The disjointness of a pair of orthants basically gives rise to one inequality constraint.  Naively one could express each equality as the conjunction of two inequalities, but this is not good enough to make the reduction work for size-4 independent set
(where we have only ${4\choose 2}=6$ inequality constraints in total).
We initially tried reducing from $C_4$ in sparse graphs, unsuccessfully (we
can only obtain a lower bound for size-5 independent set this way),
but as it turns out, reducing from hypercliques in 3-uniform hypergraphs
allows for a more symmetric and elegant proof:

\begin{theorem}\label{thm:boxes:I4:lb}
If there is an algorithm for finding a size-$4$ independent set
of $n$ orthants in $\R^6$ in $O(n^{4/3-\eps})$ time for some constant $\eps>0$,
then there is an algorithm for finding a size-$4$ hyperclique in
a $3$-uniform hypergraph with $N$ vertices in $O(N^{4-\eps'})$ time
for some constant $\eps'>0$.
\end{theorem}
\begin{pproof}
By a standard color-coding technique~\cite{AlonYZ95}, it suffices to consider the 4-hyperclique problem for a 4-partite 3-uniform hypergraph $G=(V,E)$.  In other words, $V$ is divided into four
sets $X,Y,Z,W$, and
we seek four vertices $x\in X$, $y\in Y$, $z\in Z$, and $w\in W$
such that $xyz,yzw,zwx,wxy\in E$.

First, by relabeling, assume that the vertices in $X$ are in $\{1,2,\ldots,N\}$, the vertices in $Y$ are in $\{U,2U,$ $\ldots,NU\}$,
the vertices in $Z$ are in $\{U^2,2U^2,\ldots,NU^2\}$, and
the vertices in $W$ are in $\{U^3,2U^3,\ldots,NU^3\}$, for a sufficiently large integer $U$ (say, $U=N+1$).

Create the following orthants in $\R^6$ for 
each $xy'z'',yz'w'',zw'x'',wx'y''\in E$, where $x,x',x''\in X$,
$y,y',y''\in Y$, $z,z',z''\in Z$, and $w,w',w''\in W$:

\newcommand{\timess}{\!\!\!\!\!\times\!\!\!\!\!}
\[ \!\!\begin{array}{rccccccccccc}
\alpha_{xy'z''} =\!\!\!\!& (-\infty,y'\!+\!z'') &\timess& \R &\timess& \R &\timess& (x\!+\!y',\infty) &\timess& (-\infty,x\!-\!z'') &\timess& \R\\
\beta_{yz'w''} =\!\!\!\!& (y\!+\!z',\infty) &\timess& (-\infty,z'\!+\!w'') &\timess& \R &\timess& \R &\timess& \R &\timess& (-\infty,y\!-\!w'')\\
\gamma_{zw'x''} =\!\!\!\!& \R &\timess& (z\!+\!w',\infty) &\timess& (-\infty,w'\!+\!x'') &\timess& \R &\timess& (x''\!-\!z,\infty) &\timess& \R\\
\delta_{wx'y''} =\!\!\!\!& \R &\timess& \R &\timess& (w\!+\!x',\infty) &\timess& (-\infty,x'\!+\!y'') &\timess& \R &\timess & (y''\!-\!w,\infty)
\end{array}
\]
(Open intervals are used here, but
they can be avoided easily by slight perturbations of the
coordinate values.)
We solve the independent set problem on this
set $S$ of $O(N^3)$ orthants.

To see the correctness of this reduction, 
suppose that $G$ has a 4-hyperclique $\{x,y,z,w\}$
with $x\in X$, $y\in Y$, $z\in Z$, and $w\in W$.
Then clearly $\{\alpha_{xyz},\beta_{yzw},\gamma_{zwx},\delta_{wxy}\}$
is a size-4 independent set in $S$.

The other direction is more interesting.  Suppose $S$ has 
a size-4 independent set.  Then it must be of the form
$\{\alpha_{xy'z''},\beta_{yz'w''},\gamma_{zw'x''},\delta_{wx'y''}\}$
with $xy'z'',yz'w'',zw'x'',wx',y''\in E$ and
$x\in X$, $y\in Y$, $z\in Z$, and $w\in W$ (since it cannot contain
two $\alpha$'s, nor two $\beta$'s, etc.).
The pairwise disjointness of these four orthants implies the following
6 constraints:
\[
\begin{array}{lllllll}
y'+z'' &\le& y+z' &\qquad&  x-z'' &\le & x''-z\\
z'+w'' &\le& z+w' &&        y-w'' &\le& y''-w\\
w'+x'' &\le& w+x'\\
x'+y'' &\le& x+y'
\end{array}
\]
Summing the left 4 constraints
yields $x''+y''+w''+z''\le x+y+z+w$.
On the other hand,
summing the right 2 constraints
yields $x+y+z+w\le x''+y''+w''+z''$.
Thus, we must have equality on \emph{all} of the constraints.
On the other hand, $y'+z''=y+z'$ implies
$y'=y$ and $z''=z'$, since $y',y\in \{U,2U,\ldots,NU\}$
and $z',z''\in \{U^2,2U^2,\ldots,NU^2\}$.
Repeating this argument gives us $x=x'=x''$, $y=y'=y''$,
$z=z'=z''$, and $w=w'=w''$.
Thus, $xyz,yzw,zwx,wxy\in E$, i.e., $\{x,y,z,w\}$ is a 4-hyperclique in~ $G$.

Hence, if the size-4 independent set problem for orthants could be solved in $O(n^{4/3-\eps})$ time,
then the 4-hyperclique problem could be solved in $O((N^3)^{4/3-\eps})=
O(N^{4-3\eps})$ time.
\end{pproof}

\section{Preliminaries}

From this section onward, we turn to upper bounds, i.e., algorithms.
As mentioned, we focus on the detection version of the problems.

Some of our algorithms can solve the colored
version of the problems, where the geometric objects are $k$-colored
and we seek a $k$-chromatic $k$-vertex pattern.  For example, for $C_k$,
we seek a $k$-cycle that has a vertex of each of the $k$ colors.
By a standard \emph{color-coding} technique~\cite{AlonYZ95},
the original version reduces to the $k$-colored version while increasing
the running time by only an $\OO(1)$ factor (assuming that $k$ is a constant).
Thus, all results on colored problems are automatically applicable to the original monochromatic problems, but not necessarily vice versa.

Some of our results also hold in a more general setting than
intersection graphs.  We introduce the following definition:

\begin{definition}
A \emph{$k$-partite range graph} is an undirected graph $G=(V,E)$,
where $V$ is divided into $k$ parts $V^{(1)},\ldots,V^{(k)}$, and
for each $\ell,\ell'\in\{1,\ldots,k\}$ with $\ell\neq\ell'$,
each vertex $u\in V^{(\ell)}$ is associated with a point $p^{(\ell, \ell')}(u)$
and each vertex $v\in V^{(\ell')}$ is associated with a geometric range $R^{(\ell,\ell')}(v)$,
such that $uv\in E$ iff $p^{(\ell,\ell')}(u)\in R^{(\ell,\ell')}(v)$.
\end{definition}

Note that since $G$ is undirected, the above definition implicitly requires that
$p^{(\ell,\ell')}(u)\in R^{(\ell,\ell')}(v)$ iff
$p^{(\ell',\ell)}(v)\in R^{(\ell',\ell)}(u)$.

Intersection graphs of boxes can be realized as range graphs,
since we can map a box $u = [x_1,x_1']\times \cdots\times [x_d,x_d']\subset \R^d$
to a point $p(u)=(x_1,x_1',\ldots,x_d,x_d')\in \R^{2d}$,
and another box $v= [a_1,a_1']\times \cdots\times [a_d,a_d']\subset \R^d$
to a range $R(v)=\{(x_1,x_1',\ldots,x_d,x_d')\in \R^{2d} :
\forall i,\ (a_i\le x_i\le a_i')\ \vee\ (a_i\le x_i'\le a_i')\ 
\vee\ (x_i\le a_i\le x_i')\}$.  These ranges are all unions
of $O(1)$ boxes in $2d$ dimensions.
Similarly, intersection graphs of axis-aligned orthogonal polyhedra of $O(1)$ size
can be realized as range graphs, where the ranges are unions of
$O(1)$ boxes in a larger constant dimension.
The complements of these intersection graphs are also range graphs.

Intersection graphs of 2D line segments can also be realized as range graphs,
since we can map a line segment $u$ with endpoints $(x,y)$ and $(x',y')$,
slope $\xi$, and intercept $\eta$
to a point $p(u)=(x,y,x',y',\xi,\eta)\in \R^6$,
and we can map another line segment $v$ with endpoints $(a,b)$
and $(a',b')$ 
to a range $R(v)$ consisting of all points $(x,y,x',y',\xi,\eta)\in\R^6$
such that $(x,y)$ and $(x',y')$ are on different sides of the line
through $(a,b)$ and $(a',b')$, and 
$(a,b)$ and $(a',b')$ are on different sides of the line
with slope $\xi$ and intercept $\eta$.  These ranges are polyhedral
regions of $O(1)$ complexity.  Although these regions reside in 6D,
they actually have ``intrinsic'' dimension~2, in the following sense:

\begin{definition}
Recall that a \emph{semialgebraic set} $\gamma$ is the set of all points 
$(x_1,\ldots,x_d)\in\R^d$ satisfying a Boolean combination of some finite
collection of polynomial inequalities in $x_1,\ldots,x_d$.  
We say that $\gamma$ has \emph{intrinsic dimension $d'$}
if each of these polynomial inequalities depends on at most $d'$
of the $d$ variables $x_1,\ldots,x_d$.
\end{definition}

Similarly, intersection graphs of 2D triangles or polygons of $O(1)$ size
can be realized as range graphs, where the ranges are polyhedral
regions of $O(1)$ complexity, dimension $O(1)$, and intrinsic dimension~2.
The complements of these intersection graphs are also range graphs.

Note that algorithms for finding $k$-chromatic $K_k$ in $k$-partite range graphs
can automatically be used to find $k$-chromatic induced or non-induced $X_k$ for any $k$-vertex subgraph $X_k$ for constant $k$:
in $k$-chromatic $K_k$, we seek $k$ vertices $v_1^*\in V^{(1)},\ldots,v_k^*\in V^{(k)}$ such that $v_\ell^*v_{\ell'}^*$ is in the graph for every pair
$(\ell,\ell')$, but if we want $v_\ell^*v_{\ell'}^*$ to be \emph{not} present
in the graph for certain pairs $(\ell,\ell')$, we can just take the complement
of the bipartite subgraph induced by $V^{(\ell)}\cup V^{(\ell')}$,
by taking complements of the ranges.  (The same cannot be said for
algorithms that are specialized to intersection graphs.)

In all our results concerning intersection graphs,
we assume that we are given the geometric representation, i.e.,
the geometric objects.  For range graphs,
we are given the points $p^{(\ell\ell')}(u)$ and ranges
$R^{(\ell\ell')}(v)$.

\IGNORE{

\begin{definition}
Let $V_1,\ldots,V_k$ be disjoint sets.
Suppose that for each $i,j$ and each $u\in V_i$,
we are given a point $p_ a geometric object/range $R_{ij}(u)\subset\R^d$,
satisfying the properties that $R_{ii}(u)=\emptyset$, and
$v\in R_{ij}(u)$ iff $u\in R_{ji}(v)$.
We define the \emph{$k$-partite range graph} of $(V_1,\ldots,V_k)$ to be
$G=(V_1\cup\cdots\cup V_k, E)$ where for each $u\in V_i$ and
$v\in V_j$, we have $uv\in E$ iff
$v\in R_{ij}(u)$.
\end{definition}

\begin{definition}
Let $\Range$ be a family of geometric objects (ranges) in $\R^d$.
A  $k$-partite \emph{$\Range$-range graph} is a graph of the
form $G=(V_1\cup \cdots\cup V_k, E)$, where $V_1,\ldots,V_k$ are point sets in $\R^d$, and for each $u\in V_i$ and $v\in V_j$,
we have $uv\in E$
iff $u\in R^{(ij)}(v)$ for some range $R^{(ij)}(v)\in\Range$,
with $R^{(ii)}(v)=\emptyset$.

A $k$-partite \emph{$d$-dimensional orthogonal-range graph} of size~$b$ refers
to a $k$-partite $\Range$-range graph where $\Range$ is the family of all axis-aligned orthogonal polyhedra with at most $b$ faces.
\end{definition}
}

\section{Technique 1: Biclique Covers}\label{sec:bicliques}

Our first technique involves the notion of biclique covers, which
provides a compact representation of graphs and has found many
geometric applications (e.g., see \cite{AgarwalAAS94,Chan06}):

\begin{definition}
Given a graph $G=(V,E)$, a \emph{biclique cover} is 
a collection of pairs of subsets $\{(A_1,B_1),$ $\ldots,(A_s,B_s)\}$, such that $E=\bigcup_{i=1}^s (A_i\times B_i)$.
The \emph{size} of the cover refers to
$\sum_{i=1}^s (|A_i|+|B_i|)$.
\end{definition}

For range graphs where the ranges are orthogonal objects (union of $O(1)$ boxes), it is known that
there exist biclique covers of near linear size, by techniques for orthogonal range searching, namely, \emph{range trees}~\cite{AgaEriSURV,BerBOOK}.  (For those familiar with range searching,
we think of the right part of the graph as the input
point set and the left part as the queries;
the $B_i$'s are the ``canonical subsets'' of the data structure, and $A_i$
corresponds to the set of all queries whose answers involve the canonical subset $B_i$.)

\begin{lemma}\label{lem:bicliques:boxes}
Consider a bipartite range graph $G=(V,E)$ with $n$ vertices, where where each range is a union of $O(1)$ boxes in a constant dimension.  
We can compute a biclique cover $\{(A_1,B_1),\ldots,(A_s,B_s)\}$ 
of $\OO(n)$ size 
in $\OO(n)$ time.  Furthermore, each vertex appears in $\OO(1)$ subsets $A_i$
and $\OO(1)$ subsets $B_i$.
\end{lemma}

For range graphs where the ranges have intrinsic dimension~2,
the following lemmas provide two bounds on biclique covers: the
first is lopsided (the $A_i$'s are sparser than the $B_i$'s) and the second is balanced.
All but one of our applications will use the first.
Both follow from standard techniques on 2D (nonorthogonal) range
searching, namely, \emph{multi-level cutting trees} \cite{Clarkson87,AgaEriSURV,Matousek94survey,Agarwal17survey}.  For the sake of completeness, we include quick proofs in
Appendix \ref{app:bicliques:2d:0}--\ref{app:bicliques:2d}.

\begin{lemma}\label{lem:bicliques:2d:0}
Consider a bipartite range graph $G=(V,E)$ with $n$ vertices, where
each range is a semialgebraic set of constant description complexity
in a constant dimension, with intrinsic dimension~$2$.  
We can compute a biclique cover $\{(A_1,B_1),\ldots,(A_s,B_s)\}$ 
of $\OOO(n^{3/2})$ size
in $\OOO(n^{3/2})$ time. 
Furthermore, each vertex appears in $\OO(1)$ subsets $A_i$,
and for any $r$, the number of subsets $B_i$ of size $\Theta(n/r)$ is $\OOO(r^2)$.
\end{lemma}

\begin{lemma}\label{lem:bicliques:2d}
Consider a bipartite range graph $G=(V,E)$ with $n$ vertices, where
each range is a semialgebraic set of constant description complexity 
in a constant dimension, with intrinsic dimension~$2$.  
We can compute a biclique cover $\{(A_1,B_1),\ldots,(A_s,B_s)\}$ 
of $\OOO(n^{4/3})$ size
in $\OOO(n^{4/3})$ time. 
Furthermore, for any~$r$, the number of subsets $A_i$ and $B_i$ of size
$\Theta(n/r)$ is $\OOO(r^{4/3})$. 
\end{lemma}

Generally, intersection graphs for line segments require biclique cover
size $\OOO(n^{4/3})$.  But if the graph is known to be $K_k$-free, we observe
that near-linear size is possible by adapting known techniques
based on \emph{segment trees}~\cite{ChazelleEGS94}.  We include a proof in
Appendix~\ref{app:bicliques:segs}.


\begin{lemma}\label{lem:bicliques:segs}
Let $k\ge 3$ be a constant.
Consider the intersection graph $G$ of $n$ line segments
in $\R^2$. 
We can either find a $k$-clique in $G$, or
compute a biclique cover $\{(A_1,B_1),\ldots,(A_s,B_s)\}$ of $\OO(n)$ size, 
in $\OO(n)$ time. 
Furthermore, in this cover, 
each element appears in $\OO(1)$ subsets $A_i$ and $\OO(1)$ subsets $B_i$.
\end{lemma}

\ssubsection{$C_k$ in box range graphs}

As a direct application of biclique covers, we obtain the
following result on finding $k$-cycles in box range graphs,
and thus, intersection graphs of boxes.  Note that the most
obvious way to sparsify a graph using biclique covers is to create
a new vertex per biclique (as was done, e.g., in~\cite{FederM95,Chan06}),
but this would reduce our problem to finding $C_{2k}$ in sparse graphs, which is more expensive than finding $C_k$.

\begin{theorem}\label{thm:boxes:Ck}
Let $k\ge 3$ be a constant.
Consider a $k$-partite range graph $G=(V,E)$ with $n$ vertices, where each range is a union of $O(1)$ boxes in a constant dimension.  We can find a $k$-chromatic $k$-cycle in $G$ in $\OO(T_k(n))$ time, where $T_k(m)$ denotes
the time complexity for finding a $k$-cycle in a sparse directed graph with $m$ edges.
\end{theorem}
\begin{pproof}
Assume that $V$ is divided into parts $V^{(1)},\ldots,V^{(k)}$.
We seek a $k$-cycle $v_1^*\cdots v_k^*$ with $v_1^*\in V^{(1)}$, \ldots, $v_k^*\in V^{(k)}$ (since all $k!$ cases can be handled similarly).

First,\footnote{
Throughout the paper, $[k]$ denotes $\{1,\ldots,k\}$.
}
 for each $\ell\in[k]$, compute a biclique cover $\{(A_1^{(\ell)},B_1^{(\ell)}),\ldots,(A_s^{(\ell)},B_s^{(\ell)})\}$ of 
$\OO(n)$ size 
for the subgraph of $G$ induced by $V^{(\ell)}\cup V^{(\ell+1)}$, by Lemma~\ref{lem:bicliques:boxes}, where $A_i^{(\ell)}\subseteq V^{(\ell)}$ and $B_i^{(\ell)}\subseteq V^{(\ell+1)}$.  (In the superscripts, $k+1$ is considered equivalent to 1.)
Define a new directed graph $G'$:
%
\begin{itemize}
\item For each $i\in[s]$ and $\ell\in[k]$,
create a vertex $z_i^{(\ell)}$ in $G'$.
\item For each $\ell\in[k]$, for each $v\in V^{(\ell+1)}$,
for each $B_i^{(\ell)}$ containing $v$ and each $A_j^{(\ell+1)}$ containing $v$,
create an edge from $z_i^{(\ell)}$ to $z_j^{(\ell+1)}$ in $G'$.
\end{itemize}
The problem then reduces to finding a $k$-cycle in $G'$.
Since each vertex appears in $\OO(1)$ subsets $A_i^{(\ell+1)}$,
the graph $G'$ has $\OO(n)$ edges and can be constructed in $\OO(n)$ time.
\end{pproof}

According to known results on finding short cycles in sparse directed graphs:
\begin{itemize}
\item $T_3(m) = O(m^{2\omega/(\omega+1)})\le O(m^{1.408})$ by
Alon, Yuster, and Zwick~\cite{AlonYZ97};
\item $T_4(m) = O(m^{(4\omega-1)/(2\omega+1)})\le O(m^{1.478})$
and $T_5(m) = O(m^{3\omega/(\omega+2)}) \le O(m^{1.628})$
by Yuster and Zwick~\cite{YusterZ04};
\item $T_k(m) = O(m^{2-2/k})$ if $k$ is even,
and $T_k(m) = O(m^{2-2/(k+1)})$ if $k$ is odd, by Alon, Yuster, and Zwick~\cite{AlonYZ97}
(see Dalirrooyfard, Vuong, and Vassilevska W.~\cite{DalirrooyfardVW21}
for further small improvements depending on~$\omega$). 
%
%
%
\end{itemize}

\IGNORE{
\begin{corollary}
Consider a tripartite range graph $G$ with $n$ vertices, where the ranges are all axis-aligned orthogonal-polyhedral regions of constant complexity in a constant dimension.  There is an algorithm to detect a $3$-cycle in $G$ in $\OO(n^{2\omega/(\omega+1)})\le O(n^{1.408})$ time (or in $\OO(n^{4/3})$ time if $\omega=2$).
\end{corollary}
}

\ssubsection{$C_k$ in 2D segment intersection graphs}

A similar result can be obtained for intersection graphs
of line segments:

\begin{theorem}\label{thm:segs:Ck}
Let $k\ge 3$ be a constant.
Given $n$ line segments in $\R^2$, 
we can find a $k$-cycle in the intersection graph
in $\OO(T_k(n))$ time, where $T_k(m)$ denotes the
time complexity for finding a $k$-cycle in a sparse directed graph with $m$ edges.
\end{theorem}
\begin{pproof}
By color coding~\cite{AlonYZ95}, it suffices to solve a colored version
of the problem, where the input line segments have been
colored with $k$ colors, and we seek a $k$-cycle, under the assumption
that there exists a $k$-chromatic $k$-cycle.  (The $k$-cycle found
need not be $k$-chromatic.)

We follow the same approach as in the proof of Theorem~\ref{thm:boxes:Ck:even}.
The only change is that we use 
Lemma~\ref{lem:bicliques:segs} for the biclique cover computation.
If the lemma fails to compute a cover but instead finds a $K_k$,
then we have found a $k$-cycle.
\end{pproof}

Unfortunately, this approach does not directly yield
subquadratic algorithms for trichromatic $C_3$ or $I_3$
for line segments or other types of objects;
for example, for $C_3$ for line segments, we could apply the $\OOO(n^{3/2})$-size biclique cover from Lemma~\ref{lem:bicliques:2d:0}
(as we need the property that each vertex appears in $\OO(1)$ of the $A_i$'s),
but the resulting time bound would be near $(n^{3/2})^{1.408}$, which is \emph{super}quadratic!

\section{Technique 2: High-Low Tricks}\label{sec:high:low}

In this section, we describe algorithms based on combining biclique covers with the following kind of tricks: 
dividing into cases based on whether edges in the solution are ``high''
or ``low'', and at the end, choosing parameters to balance cost of the
different cases.  There are many options in how to divide into cases and
how to handle these cases efficiently, making it a challenge to find
the best time bounds (we have actually discarded many slower alternatives before arriving at the final algorithms presented here).

One interesting additional idea, used in some of our algorithms
for the low cases, is to treat a pair or tuple of objects as one
``compound'' object.  From these compound objects, we can then
apply range searching results or invoke earlier algorithms to find
a smaller pattern involving fewer objects.  Here, generalizations to
range graphs become crucial, even if we originally only care 
about intersection graphs.

\ssubsection{$C_3$ in 2D range graphs}

As a first example of the ``high-low'' approach, 
we describe an $\OOO(n^{7/4})$-time algorithm
for finding $C_3$ in range graphs with intrinsic dimension~2.
As mentioned, achieving subquadratic time is more challenging here because
the biclique cover size is larger.
To address the low case, we will use the idea of treating a pair as an object. 
Below, we will state our result in a more general unbalanced setting because
it will be needed in a later algorithm, but the reader may choose to focus on
the main case when $n'=n$.

\begin{theorem}\label{thm:segs:C3}
Consider a tripartite range graph $G=(V,E)$ where the first two parts have $O(n)$ vertices and the third part has $O(n')$ vertices, and each range is a semialgebraic set of constant description complexity in a constant dimension, with intrinsic dimension~$2$.  
We can find a $3$-cycle in $G$ in $\OOO(n(n')^{3/4})$ time, assuming that $n^{4/5}\le n'\le n^4$.
\end{theorem}
\begin{pproof}
Assume that $V$ is divided into parts $V^{(1)},V^{(2)},V^{(3)}$.
We seek a 3-cycle $C^*=v_1^*v_2^*v_3^*$ with $v_1^*\in V^{(1)}, \ldots, v_3^*\in V^{(3)}$.

First, compute a biclique cover $\{(A_1,B_1),\ldots,
(A_s,B_s)\}$ of $\OOO(n^{4/3})$ size for the subgraph of $G$ induced by $V^{(1)}\cup V^{(2)}$, by
Lemma~\ref{lem:bicliques:2d}, where $A_i\subseteq V^{(1)}$ and $B_i\subseteq V^{(2)}$.

Let $r\le n$ be a parameter.
Call an edge $uv\in E$ \emph{low} if $u\in A_i$ and $v\in B_i$
for some $i$ with $\mu_i := |A_i|+|B_i|\le n/r$, and \emph{high} otherwise.
Since the number of $i$'s with $\mu_i=\Theta(n/(2^jr))$ is
$O^*((2^jr)^{4/3})$,  
the number of low edges is 
$\OOO(\sum_{j\ge 0} (2^jr)^{4/3} (n/(2^jr))^2)
= \OOO(n^2/r^{2/3})$.
We consider two cases:%
\footnote{Of course, we don't know $C^*$ 
and so don't know in advance which case holds.
What we are saying is that if the statement for a case is true,
then the algorithm for that case is guaranteed to find an answer (though not necessarily $C^*$ itself).  By running the algorithms
for all cases, we are guaranteed to find an answer if $C^*$ exists.
If none of the cases succeeds, we can conclude that no solution exists.  A
similar comment applies to all the other algorithms in this section.}

\begin{itemize}
\item {\sc Case 1:} $v_1^*v_2^*$ is low.
For each low edge $v_1v_2$, it suffices to test whether there exists
$v_3\in V^{(3)}$ such that $v_1v_3\in E$ and $v_2v_3\in E$,
i.e., the point $(p^{(3,1)}(v_3),p^{(3,2)}(v_3))$ is in the range $R^{(3,1)}(v_1)\times R^{(3,2)}(v_2)$.
These tests reduce to $Q=\OOO(n^2/r^{2/3})$ range searching queries on $O(n')$ points in a constant dimension,
with intrinsic dimension~2.
By known range searching results \cite{Matousek92,Matousek93,AgarwalM94,AgaEriSURV,Matousek94survey,Agarwal17survey},\footnote{
Using a data structure with $\OOO(m)$ preprocessing time and $\OOO(n'/\sqrt{m})$ query time ($n'\le m\le (n')^2$)
 \cite{Matousek92,Matousek93,AgaEriSURV,Matousek94survey,Agarwal17survey}, the total time to answer $Q$ queries on $n'$ points is $\OOO(m + Qn'/\sqrt{m})$.  Choosing $m$ to balance the two terms yields the expression $\OOO((n'Q)^{2/3} + n' + Q)$.
}
these queries can be answered
in time $\OOO((n'Q)^{2/3} + n' + Q)=\OOO(n^{4/3}(n')^{2/3}/r^{4/9} + n' + n^2/r^{2/3})$.

\item {\sc Case 2:} $v_1^*v_2^*$ is high.
Guess the index $i$ with $\mu_i > n/r$, such that
$(v_1^*,v_2^*)\in A_i\times B_i$.
There are $\OOO(r^{4/3})$ choices for $i$.
For each $v_3\in V^{(3)}$, we test whether there exists
$v_1\in A_i$ such that $v_1v_3\in E$, i.e., $v_1$ is in the
range $R^{(3,1)}(v_3)$, and similarly test whether 
there exists
$v_2\in B_i$ such that $v_2v_3\in E$, i.e., $v_2$ is in the
range $R^{(3,2)}(v_3)$.  If both tests return true, we have found 
a 3-cycle $v_1v_2v_3$ (since there is an edge between any $v_1\in A_i$
and any $v_2\in B_i$).
These tests reduce to $O(n')$ range searching queries on 
$O(\mu_i)$ points, with intrinsic dimension~2.
By known range searching results, these queries can be answered in
$\OOO((n'\mu_i)^{2/3} + n' + \mu_i)$ time.
Recall that the number of $i$'s with $\mu_i=\Theta(2^jn/r)$ is $\OOO((r/2^j)^{4/3})$.
Summing over all $i$, we get the total time bound
$\OOO(\sum_{j\ge 0} (r/2^j)^{4/3}\cdot ((n' (2^jn/r))^{2/3} + n' + 2^jn/r))
=\OOO(n^{2/3}(n')^{2/3}r^{2/3} + n'r^{4/3} + nr^{1/3})$.
\end{itemize}

We set $r=n^{3/4}/(n')^{3/16}$ to equalize $n'r^{4/3}$ with $n^{4/3}(n')^{2/3}/r^{4/9}$.  The terms $n^{2/3}(n')^{2/3}r^{2/3}$
and $n^2/r^{2/3}$ do not dominate if $n'\ge n^{4/5}$.
\end{pproof}

\ssubsection{$I_3$ in 2D segment intersection graphs}

The preceding algorithm can immediately be used to find size-3 independent
sets for line-segment intersection graphs, because
the complement graph can be realized as a range graph with intrinsic
dimension~2.  But in this subsection, we will present a different 
high-low division that yields a faster algorithm.  It is interesting that
already for $k=3$, there are very different ways to apply the high-low trick.

We first note in the following lemma that size-2 independent sets are easy to find for
line segments: 

\begin{lemma}\label{lem:segs:I2}
Given $n$ red/blue line segments in $\R^2$, we can find a bichromatic pair of nonintersecting line segments
in $\OO(n)$ time. 
\end{lemma}
\begin{pproof}
We seek a red segment $s_1^*$ and a blue segment $s_2^*$ that do not intersect.  There are 4 possibilities:
(i)~$s_1^*$ is above the line through $s_2^*$,
(ii)~$s_1^*$ is below the line through $s_2^*$,
(iii)~$s_2^*$ is above the line through $s_1^*$, or
(iv)~$s_2^*$ is above the line through $s_1^*$.
It suffices to consider case~(iv), since the other cases can be handled similarly.

Let $p_1^*$ and $q_1^*$ be the left and right endpoints of $s_1^*$.
Case~(i) can be further broken into two subcases:
(a)~$p_1^*$ is above the line through $s_2^*$ and the slope of $s_1^*$ is greater than the slope of $s_2^*$, or
(b)~$q_1^*$ is above the line through $s_2^*$ and the slope of $s_1^*$ is less than the slope of $s_2^*$.
It suffices to consider case~(a).

\newcommand{\HH}{{\cal H}}
To this end, we maintain the upper hull $\HH_\mu$ of the set of 
the left endpoints $p_1$ of all red segments $s_1$ with slope greater than $\mu$, as $\mu$ decreases from $\infty$ to $-\infty$.
The hull undergoes $O(n)$ insertions of points, and can be
maintained in $O(\log n)$ time per insertion by a known incremental convex hull algorithm~\cite{Preparata79}.  For each blue segment $s_2$ with slope $m$, we just check whether there exists a vertex $p_1$ of $\HH_m$ that is above the line through $s_2$, in $O(\log n)$ time by binary search.  The total running time is $O(n\log n)$.
\end{pproof}

\begin{theorem}\label{thm:segs:I3}
Given $n$ red/blue/green line segments in $\R^2$,
we can find a trichromatic size-$3$ independent set in 
$\OO(n^{4\omega/(2\omega+1)})\le O(n^{1.652})$ time.
\end{theorem}
\begin{pproof}
Let $G=(V,E)$ denote the complement of the intersection graph.
Let $V^{(1)},V^{(2)},V^{(3)}$ be the color classes.
We seek a 3-cycle $C^*=v_1^*v_2^*v_3^*$ in $G$ with $v_1^*\in V^{(1)}$, \ldots, $v_3^*\in V^{(3)}$.

First, for each $\ell\in [3]$, compute a biclique cover $\{(A_1^{(\ell)},B_1^{(\ell)}),\ldots,
(A_s^{(\ell)},B_s^{(\ell)})\}$ of $\OOO(n^{3/2})$ size for the subgraph of $G$ induced by $V^{(\ell)}\cup V^{(\ell+1)}$, by
Lemma~\ref{lem:bicliques:2d:0}.  (The ``lopsided'' form of biclique cover
turns out to be better here.  In the superscripts, 4 is considered equivalent to 1.)

Let $r\le n$ be a parameter.
Call an edge $uv\in E$ \emph{low} if $u\in A_i^{(\ell)}$ and $v\in B_i^{(\ell)}$
for some $i$ with $|B_i^{(\ell)}|\le n/r$, and \emph{high} otherwise. 
The number of low edges is 
$\OO(n^2/r)$, since each vertex appears in $\OO(1)$ subsets $A_i^{(\ell)}$.
We consider two cases:

\begin{itemize}
\item {\sc Case 1:} At least one edge of $C^*$ is low.
Say it is $v_1^*v_2^*$.
First, for each $i$, compute the subset $X_i = \{v_1\in V^{(1)}:
\mbox{$v_1v_2$ is a low edge and $v_2\in A_i^{(2)}$}\}$.
Since each vertex appears in $\OO(1)$ subsets $A_i^{(2)}$,
we have $\sum_i |X_i|=\OO(n^2/r)$ and the $X_i$'s can be
generated in $\OO(n^2/r)$ time.

Guess the index $i$ such that $(v_2^*,v_3^*)\in A_i^{(2)}\times B_i^{(2)}$.  We check whether there exists an edge
between $X_i$ and $B_i^{(2)}$.  By Lemma~\ref{lem:segs:I2}, this takes
$\OO(|X_i|+|B_i^{(2)}|)$ time.  If such an edge is found between $v_1\in X_i$ 
and $v_3\in B_i^{(2)}$,  then since $v_1$ is adjacent to some $v_2\in A_i^{(2)}$, we have found a 3-cycle $v_1v_2v_3$.
Summing over all $i$, we get the total time bound
$\OOO(n^2/r + n^{3/2})$.
\item {\sc Case 2:} All three edges of $C^*$ are high.
We follow the approach in the proof of Theorem~\ref{thm:boxes:Ck}.  Define a new directed graph $G'$:
\begin{itemize}
\item For each $i\in[s]$ and $\ell\in[3]$ with $|B_i^{(\ell)}|>n/r$,
create a vertex $z_i^{(\ell)}$ in $G'$.
\item For each $v\in V$ and $\ell\in[3]$,
for each $B_i^{(\ell)}$ containing $v$ and each $A_j^{(\ell+1)}$ containing $v$, if $z_i^{(\ell)}$ and $z_j^{(\ell+1)}$ exist,
create an edge from $z_i^{(\ell)}$ to $z_j^{(\ell+1)}$ in $G'$.
\end{itemize}
The problem then reduces to finding a 3-cycle in $G'$.
Since each vertex appears in $\OO(1)$ subsets $A_j^{(\ell+1)}$,
the graph $G'$ can be constructed in time
$\OO(\sum_{i,\ell:\,|B_i^{(\ell)}|>n/r}|B_i^{(\ell)}|)
=\OOO(\sum_{j\ge 0} (r/2^j)^2 \cdot (2^jn/r)) = \OOO(nr)$
(recalling that the number of $i$'s with $|B_i^{(\ell)}|=\Theta(2^jn/r)$
is $O((r/2^j)^2)$).
Since $G'$ has $\OO(r^2)$ vertices,
we can detect a 3-cycle in $G'$ by matrix multiplication
in $\OO((r^2)^{\omega})$ time.
\end{itemize}

We set $r=n^{2/(2\omega+1)}$ to equalize $r^{2\omega}$
with $n^2/r$.
\end{pproof}

\ssubsection{$K_4$ (and $K_k$) in box range graphs}

We now present a subquadratic algorithm for finding 4-cliques for boxes.
The details will now get more elaborate.  The key idea in the low case
is to treat a pair as an object so as to reduce to a $C_3$ problem.
The high case has many options; the best way we have come up with is to guess away two edges of $K_4$ so as
to reduce to $C_4$ subproblems.

First, as subroutines, we need variants of the known $\OO(m^{1.408})$-time
algorithm for $C_3$ and a known $\OO(m^{3/2})$-time
algorithm for $C_4$ in a sparse graph by Alon, Yuster, and Zwick~\cite{AlonYZ97}, when the graph is 3- or 4-partite 
and is lopsided, as stated in the two lemmas below.  The generalizations are straightforward and
are shown in Appendix~\ref{app:sparse:C3}--\ref{app:sparse:C4}.
(Note that when $m=m'$, the expression $\min_\D (m\D + M(m/\D,m/\D,m/\D))$
indeed recovers the $O(m^{1.408})$ bound under the current matrix multiplication
exponent.)

\begin{lemma}\label{lem:sparse:C3}
Consider a tripartite graph $G$, where the vertices are divided into parts $V^{(1)},V^{(2)},V^{(3)}$, and there are $O(m)$ edges between $V^{(1)}$ and $V^{(2)}$, $O(m)$ edges between $V^{(2)}$ and $V^{(3)}$, and $O(m')$ edges between $V^{(3)}$ and $V^{(1)}$.  Then we can find a $3$-cycle in $G$ in
$O(m'\D + M(m/\D,m^2/(\D m'),m/\D))$ time for any given~$\D$,
where $M(n_1,n_2,n_3)$ is the complexity of multiplying an $n_1\times n_2$ with an $n_2\times n_3$ matrix.
\end{lemma}

\begin{lemma}\label{lem:sparse:C4}
Consider a $4$-partite graph $G$, where the vertices are divided into parts $V^{(1)},V^{(2)},V^{(3)},V^{(4)}$, and there are $O(m)$ edges between $V^{(1)}$ and $V^{(2)}$, $O(m')$ edges between $V^{(2)}$ and $V^{(3)}$, $O(m)$ edges between $V^{(3)}$ and $V^{(4)}$, and $O(m')$ edges between $V^{(4)}$ and $V^{(1)}$.  Then we can find a $4$-chromatic $4$-cycle in $G$ in
$O(m\sqrt{m'} + m'\sqrt{m})$ time.
\end{lemma}

\begin{theorem}\label{thm:boxes:K4}
Consider a $4$-partite range graph $G=(V,E)$ with $n$ vertices, 
where each range is a union of $O(1)$ boxes in a constant dimension.  We can find a $4$-clique in $G$ in $O(n^{1.715})$ time (or in $\OO(n^{12/7})$ time if $\omega=2$).
\end{theorem}
\begin{pproof}
Assume that $V$ is divided into parts $V^{(1)},\ldots,V^{(4)}$.
We seek a 4-clique $K^*=\{v_1^*,\ldots,v_4^*\}$ with $v_1^*\in V^{(1)}$, \ldots, $v_4^*\in V^{(4)}$.

First, for each $\ell,\ell'\in [4]$ with $\ell\neq\ell'$, compute a biclique cover $\{(A_1^{(\ell,\ell')},B_1^{(\ell,\ell')}),\ldots,(A_s^{(\ell,\ell')},B_s^{(\ell,\ell')})\}$
of $\OO(n)$ size for the subgraph of $G$ induced by $V^{(\ell)}\cup V^{(\ell')}$, by
Lemma~\ref{lem:bicliques:boxes}, where $A_i^{(\ell,\ell')}\subseteq V^{(\ell)}$ and $B_i^{(\ell,\ell')}\subseteq V^{(\ell')}$.

Let $r\le n$ be a parameter.
Call an edge $uv\in E$ \emph{low} if $u\in A_i^{(\ell,\ell')}$ and $v\in B_i^{(\ell,\ell')}$
for some $i$ with $\mu_i^{(\ell,\ell')} := |A_i^{(\ell,\ell')}|+|B_i^{(\ell,\ell')}|\le n/r$, and \emph{high} otherwise. 
The number of low edges is $\OO(n^2/r)$.
We consider two cases:

\begin{itemize}
\item {\sc Case 1:} At least one edge of $K^*$ is low.  Say it is $v_1^*v_2^*$.  Define a tripartite graph $\GG$:
\begin{itemize}
\item Each low edge $v_1v_2$ with $v_1\in V^{(1)}$ and $v_2\in V^{(2)}$ is a vertex in $G'$.  Each vertex in $V^{(3)}$ is a vertex in $\GG$.  Each vertex in $V^{(4)}$ is a vertex in $\GG$.
\item Create an edge between $v_1v_2$ and $v_3\in V^{(3)}$ in $\GG$
iff $v_1v_3\in E$ and $v_2v_3\in E$.  Create an edge between $v_1v_2$ and $v_4\in V^{(4)}$ in $\GG$
iff $v_1v_4\in E$ and $v_2v_4\in E$.  Create an edge between
$v_3\in V^{(3)}$ and $v_4\in V^{(4)}$  in $\GG$ iff $v_3v_4\in E$.
\end{itemize}
The problem then reduces to finding a 3-cycle in $\GG$.
Note that $\GG$ can be realized as a range graph,
since the condition ``$v_1v_3\in E$ and $v_2v_3\in E$'' is true
iff the point $(p^{(1,3)}(v_1),p^{(2,3)}(v_2))$ lies in the range
$R^{(1,3)}(v_3)\times R^{(2,3)}(v_3)$,
iff the point $(p^{(3,1)}(v_3),p^{(3,2)}(v_3))$ lies in the range
$R^{(3,1)}(v_1)\times R^{(3,2)}(v_2)$.
The condition ``$v_1v_4\in E$ and $v_2v_4\in E$'' is similar.
These ranges are all unions of $O(1)$ boxes in some constant dimension. 
We can thus follow the approach in the proof of Theorem~\ref{thm:boxes:Ck} to reduce 3-cycle detection in $\GG$
to 3-cycle detection in some new graph $G'$.
Now, $\GG$ is a tripartite graph, where the first part has $\OO(n^2/r)$
vertices and the second and third parts have $O(n)$ vertices.
In this unbalanced scenario, the graph $G'$ constructed 
in the proof of Theorem~\ref{thm:boxes:Ck} is tripartite,
where there are $m=O(n)$ edges between the first and the second part and between the second and third part, and $m'=\OO(n^2/r)$ edges between the third and first part.
By Lemma~\ref{lem:sparse:C3}, we can solve the problem
in $O(n^2\D/r + M(n/\D,r/\D,n/\D))$ time.

\item {\sc Case 2:} All edges of $K^*$ are high.  In particular, $v_1^*v_3^*$ and $v_2^*v_4^*$ are high.
Guess the index $i$ with $\mu_i^{(1,3)} > n/r$,
such that $(v_1^*,v_3^*)\in A_i^{(1,3)}\times B_i^{(1,3)}$.
Guess the index $j$ with $\mu_j^{(2,4)} > n/r$,
such that $(v_2^*,v_4^*)\in A_j^{(2,4)}\times B_j^{(2,4)}$.
There are $\OO(r)$ choices for $i$ and $\OO(r)$ choices for $j$.
It suffices to solve the 4-clique detection problem in the
subgraph induced by $A_i^{(1,3)} \cup
A_j^{(2,4)}\cup B_i^{(1,3)} \cup B_j^{(2,4)}$,
but since all vertices in $A_i^{(1,3)}$ are adjacent to
all vertices in $B_i^{(1,3)}$ and all vertices in $A_j^{(2,4)}$ are adjacent to
all vertices in $B_j^{(2,4)}$, the problem reduces to detecting a 4-chromatic 4-cycle in this subgraph.
We can follow the approach in the proof of Theorem~\ref{thm:boxes:Ck} to reduce 
to 4-cycle detection in a new sparse graph, where there
are $O(\mu_i^{(1,3)})$ edges between the first and second part and
between the third and fourth part, and $O(\mu_j^{(2,4)})$ edges
between the second and third part and between the fourth
and first part.
By Lemma~\ref{lem:sparse:C4}, we can solve the problem in
$O(\mu_i^{(1,3)}\sqrt{\mu_j^{(2,4)}} + \mu_j^{(2,4)}\sqrt{\mu_i^{(1,3)}})$ time.

Since $\sum_j \mu_j^{(2,4)}=\OO(n)$ and there are $\OO(r)$ choices for $j$, we have
$\sum_{j} \sqrt{\mu_j^{(2,4)}} = \OO(\sqrt{nr})$.
Since $\sum_i \mu_i^{(1,3)}=\OO(n)$, we have
$\sum_{i}\sum_j \mu_i^{(1,3)}\sqrt{\mu_j^{(2,4)}} = \OO(n^{3/2}r^{1/2})$.
We can sum $\mu_j^{(2,4)}\sqrt{\mu_i^{(1,3)}}$ similarly.  Thus, the total running time 
is $\OO(n^{3/2}r^{1/2})$.
\end{itemize}

We set $\D=r^{3/2}/n^{1/2}$ to equalize $n^2\D/r$ with $n^{3/2}r^{1/2}$.  The overall running time is 
\[ \OO(n^{3/2}r^{1/2} + M((n/r)^{3/2}, (n/r)^{1/2}, (n/r)^{3/2}))\ =\ \OO(n^{3/2}r^{1/2} + (n/r)^{3\beta/2}),
\] where
$\beta=\omega(1,1/3,1)< 2.002$~\cite{GallU18}.  Setting $r=n^{(3\beta-3)/(3\beta+1)}$
gives the bound $\OO(n^{6\beta/(3\beta+1)}) \le O(n^{1.715})$.
\end{pproof}

From this algorithm for $K_4$, we immediately get a result
for $K_k$ for larger $k$ by treating $(k/4)$-tuples of objects
as one compound object (which we can do for range graphs):

\begin{corollary}\label{cor:boxes:Kk}
Let $k$ be a constant integer divisible by $4$.
Consider a $k$-partite range graph $G=(V,E)$ with $n$ vertices, where each range is a union of $O(1)$ boxes in a constant dimension.  We can find a $k$-clique in $G$ in $O(n^{0.429k})$ time.
\end{corollary}
\begin{pproof}
Assume that $V$ is divided into parts $V^{(1)},\ldots,V^{(k)}$.

Define a 4-partite graph $\GG$:
\newcommand{\VV}{\widehat{V}}
\begin{itemize}
\item The vertices are $\VV^{(1)}\cup\cdots\cup \VV^{(4)}$ 
where $\VV^{(\ell)}$ is the set of all $(k/4)$-cliques in 
the subgraph of $G$ induced by
$V^{(\ell k/4 + 1)}\times\cdots\times V^{((\ell+1)k/4)}$.
\item For each $\ell,\ell'\in [4]$ with $\ell\neq\ell'$, create an edge between $\{v_{\ell k/4+1},\ldots,v_{(\ell+1)k/4}\}\in \VV^{(\ell)}$ and 
$\{v_{\ell' k/4+1},\ldots,v_{(\ell'+1)k/4}\}\in  \VV^{(\ell')}$ iff
$v_{\ell k/4+i}v_{\ell'k/4+j}\in E$ for all $i,j\in[k/4]$.
\end{itemize}
Note that $\GG$ is a range graph with $O(n^{k/4})$ vertices, since
$v_{\ell k/4+i}v_{\ell'k/4+j}\in E$ for all $i,j\in [k/4]$ iff the point $(p^{(\ell,\ell')}(v_{\ell k/4+1}),\ldots,
p^{(\ell,\ell')}(v_{\ell k/4+1}), \ldots, 
p^{(\ell,\ell')}(v_{(\ell+1)k/4}), \ldots, 
p^{(\ell,\ell')}(v_{(\ell+1)k/4}))$ lies in the range
$R^{(\ell,\ell')}(v_{\ell' k/4+1})\times \ldots\times
R^{(\ell,\ell')}(v_{(\ell'+1)k/4})\times \ldots\times
R^{(\ell,\ell')}(v_{\ell' k/4+1})\times \ldots\times
R^{(\ell,\ell')}(v_{(\ell'+1)k/4})$.
The ranges are all unions of $O(1)$ boxes in some constant dimension. 
By Theorem~\ref{thm:boxes:K4}, we can solve the problem in
$O((n^{k/4})^{1.715})$ time.
\end{pproof}

Note that if we were to start from our $O(n^{1.408})$-time
algorithm for $C_3$ in Theorem~\ref{thm:boxes:Ck}, we would
get a time bound of $O((n^{k/3})^{1.408})=O(n^{0.47k})$ for $k$
divisible by 3, which is worse.

For $k$ not divisible by 4, we can naively round up 
and get a time bound of $O(n^{1.715\lceil k/4\rceil})\le
O(n^{0.429k+O(1)})$, but improvement in the ``lower-order'' term in the
exponent should be possible with
more work.  There might be room for further small improvement over the coefficient $0.429$ itself,
perhaps by analyzing larger patterns beyond $K_4$ with more (tedious) effort.

As mentioned earlier, algorithms for $k$-chromatic $K_k$ in range graphs
can be used to find induced or non-induced copies of any $k$-vertex
subgraphs $X_k$.

\ssubsection{$K_4$ in 2D segment intersection graphs}

For $K_4$ for line-segment intersection graphs, we can
still get subquadratic time by modifying our earlier approach for $K_4$ for boxes, since there are
still near-linear biclique covers due to Lemma~\ref{lem:bicliques:segs} (at least when the graph is $K_k$ free), but the time bound increases.

First we start with a variant of Theorem~\ref{thm:segs:C3} for a special case when
near-linear biclique covers exist:

\begin{lemma}\label{lem:segs:C3:modified}
Consider a tripartite range graph $G=(V,E)$ where the first two parts have $O(n)$ vertices and the third part has $O(n')$ vertices, and each range is a semialgebraic set of constant description complexity in a constant dimension, with intrinsic dimension~$2$.  
Suppose we are given a biclique cover $\{(A_1,B_1),\ldots,
(A_s,B_s)\}$ for the subgraph induced by the first two parts with
$\sum_i (|A_i|+|B_i|)=\OO(n)$, and each vertex appears in $\OO(1)$
subsets $A_i$.
We can find a $3$-cycle in $G$ in $\OOO((nn')^{4/5})$ time, assuming that $n^{2/3}\le n'\le n^4$.
\end{lemma}
\begin{pproof}
We re-analyze the algorithm in the proof of Theorem~\ref{thm:segs:C3}, using the better biclique cover bound.
The number of low edges is now $\OO(n^2/r)$.  
\begin{itemize}
\item In Case~1, we now have $Q=\OO(n^2/r)$ and the time bound
becomes  
$\OOO((n'Q)^{2/3} + n' + Q)=\OOO(n^{4/3}(n')^{2/3}/r^{2/3} + n + n^2/r)$.
\item
In Case~2, the time bound becomes
$\OOO(\sum_{j\ge 0} (r/2^j)\cdot ((n' (2^jn/r))^{2/3} + n' + 2^jn/r))
=\OOO(n^{2/3}(n')^{2/3}r^{1/3} + n'r + n)$.
\end{itemize}

We set $r=n^{4/5}/(n')^{1/5}$ to equalize $n'r$ with $n^{4/3}(n')^{2/3}/r^{2/3}$.  The terms $n^{2/3}(n')^{2/3}r^{1/3}$
and $n^2/r$ do not dominate if $n'\ge n^{2/3}$.
\end{pproof}

\begin{theorem}\label{thm:segs:K4}
Given $n$ line segments in $\R^2$,
we can find a $4$-clique in the intersection graph in $\OOO(n^{24/13})\le O(n^{1.847})$ time.
\end{theorem}
\begin{pproof}
We modify the algorithm in the proof of Theorem~\ref{thm:boxes:K4}.
First we use 
Lemma~\ref{lem:bicliques:segs} with $k=4$ for the computation of a biclique cover of $\OO(n)$ size.
If the lemma fails to compute a cover but instead finds a $K_k$,
then we have found a 4-cycle.  Otherwise:

\begin{itemize}
\item In Case 1, we find a 3-cycle in $\GG$ using
Lemma~\ref{lem:segs:C3:modified} (with the first and third parts switched).  The time bound is now
$\OOO((n(n^2/r))^{4/5}) = \OOO(n^{12/5}/r^{4/5})$,
assuming $n^2/r \ge n^{2/3}$.
\item In Case 2, the same analysis still holds, as the biclique
cover still has $\OO(n)$ size.  The time bound remains $\OO(n^{3/2}r^{1/2})$.
\end{itemize}

We set $r=n^{9/13}$ to equalize $n^{3/2}r^{1/2}$ with $n^{12/5}/r^{4/5}$.
\end{pproof}

Note that the above theorem applies to segment intersection graphs but
not to more general range graphs, and so it cannot be applied to
find $K_k$ for larger $k$, unlike before.  To this end, we will  turn to
finding $K_6$ next in the range graph setting.

\ssubsection{$K_6$ (and $K_k$) in 2D range graphs}

To find $K_6$ in range graphs with intrinsic dimension~2,
we will handle the low case again by treating pairs as objects so as
to reduce to a $C_3$ problem, and we will handle the high case by guessing
away 9 of the 15 edges of $K_6$ so as to reduce to $C_6$ subproblems.

\begin{theorem}\label{thm:segs:K6}
Consider a $6$-partite range graph $G=(V,E)$ with $n$ vertices, where each range is a semialgebraic set of constant description complexity in a constant dimension, with intrinsic dimension~$2$. 
We can find a $6$-clique in $G$ in $\OO(n^{28(9+\omega)/(75+8\omega)})\le O(n^{3.389})$ time.
\end{theorem}
\begin{pproof}
%
%
Assume that $V$ is divided into parts $V^{(1)},\ldots,V^{(6)}$.
We seek a 6-clique $K^*=\{v_1^*,\ldots,v_6^*\}$ with $v_1^*\in V^{(1)}$, \ldots, $v_6^*\in V^{(6)}$.

First, for each $\ell,\ell'\in [6]$ with $\ell\neq\ell'$, compute a biclique cover $\{(A_1^{(\ell,\ell')},B_1^{(\ell,\ell')}),\ldots,(A_s^{(\ell,\ell')},B_s^{(\ell,\ell')})\}$ of $\OOO(n^{3/2})$ size for the subgraph of $G$ induced by $V^{(\ell)}\cup V^{(\ell')}$, by
Lemma~\ref{lem:bicliques:2d:0}, where $A_i^{(\ell,\ell')}\subseteq V^{(\ell)}$ and $B_i^{(\ell,\ell')}\subseteq V^{(\ell')}$.

Let $r\le n$ be a parameter.
Call an edge $uv\in E$ \emph{low} if $u\in A_i^{(\ell,\ell')}$ and $v\in B_i^{(\ell,\ell')}$
for some $i$ with $|B_i^{(\ell,\ell')}|\le n/r$, and \emph{high} otherwise. 
The number of low edges is $\OO(n^2/r)$.
We consider two cases:

\begin{itemize}
\item {\sc Case 1:} At least one edge of $K^*$ is low.  Say it is $v_1^*v_2^*$.  Define a tripartite graph $\GG$:
\begin{itemize}
\item Each low edge $v_1v_2$ with $v_1\in V^{(1)}$ and $v_2\in V^{(2)}$ is a vertex in $G'$.  Each edge $v_3v_4$ with $v_3\in V^{(3)}$ and $v_4\in  V^{(4)}$ is a vertex in $\GG$.  Each edge $v_5v_6$ in $v_5\in V^{(5)}$ and $v_6\in V^{(6)}$ is a vertex in $\GG$.
\item Create an edge between $v_1v_2$ and $v_3v_4$ in $\GG$ 
iff $v_1v_3,v_1v_4,v_2v_3,v_2v_4\in E$.  Create an edge between $v_1v_2$ and $v_5v_6$ in $\GG$
iff $v_1v_5,v_1v_6,v_2v_5,v_2v_6\in E$.  Create an edge between
$v_3v_4$ and $v_5v_6$ in $\GG$ iff $v_3v_5,v_3v_6,v_4v_5,v_4v_6\in E$.
\end{itemize}
Our problem then reduces to finding a 3-cycle in $\GG$.
Note that $\GG$ can be realized as a range graph, like before.
The ranges all have constant description complexity in some constant
dimension, with intrinsic dimension~2.
Now, $\GG$ is a tripartite graph, where the first part has $\OO(n^2/r)$
vertices and the second and third parts have $O(n^2)$ vertices.
By Theorem~\ref{thm:segs:C3} (with the first and third parts switched), we can solve the problem
in $\OOO(n^2 (n^2/r)^{3/4}) = \OOO(n^{7/2}/r^{3/4})$ time,
assuming that $n^2/r\ge (n^2)^{4/5}$.

\item {\sc Case 2:} All edges of $K^*$ are high.
Let $Z=\{(i,j)\in [6]^2: i<j\} - \{(1,2),(2,3),(3,4),(4,5),$ $(5,6),(1,6)\}$.
For each $(\ell,\ell')\in Z$, guess the index $j[\ell,\ell']$ with $|B_{j[\ell,\ell']}^{(\ell,\ell')}| > n/r$,
such that $(v_\ell^*,v_{\ell'}^*)\in A_{j[\ell,\ell']}^{(\ell,\ell')}\times B_{j[\ell,\ell']}^{(\ell,\ell')}$; mark all vertices in $V^{(\ell)}-A_{j[\ell,\ell']}^{(\ell,\ell')}$ and in $V^{(\ell')}-B_{j[\ell,\ell']}^{(\ell,\ell')}$
as ``invalid''.
It suffices to find a 6-chromatic 6-cycle $v_1^*\cdots v_6^*$ with $v_1^*\in V^{(1)},\ldots,v_6^*\in V^{(6)}$ in the subgraph of $G$ induced by the vertices not marked ``invalid''.
We now follow the approach in the proof of Theorem~\ref{thm:boxes:Ck}.  Define a new directed graph $G'$ (in the superscripts, 7 and 8 are considered equivalent to 1 and 2):
\begin{itemize}
\item For each $i\in[s]$ and $\ell\in[6]$ with $|B_i^{(\ell,\ell+1)}| > n/r$,
create a vertex $z_i^{(\ell)}$ in $G'$.
\item For each $\ell\in[6]$, for each $v\in V^{(\ell+1)}$ not marked ``invalid'',
for each $B_i^{(\ell,\ell+1)}$ containing $v$ and each $A_j^{(\ell+1,\ell+2)}$ containing $v$,
if $z_i^{(\ell)}$ and $z_j^{(\ell+1)}$ exist, 
create an edge from $z_i^{(\ell)}$ to $z_j^{(\ell+1)}$ in $G'$.
\end{itemize}
The problem then reduces to finding a $6$-cycle in $G'$.
Note that $G'$ has $\OOO(r^2)$ vertices.
We can detect a 6-cycle in $G'$ by matrix multiplication in $\OOO((r^2)^\omega)$ time~\cite{AlonYZ95}.

There are $\OOO(r^2)$ choices for each $j[\ell,\ell']$,
and so $\OOO((r^2)^9)$ choices overall (since $|Z|={6\choose 2}-6=9$).
The total time is $\OOO((r^2)^9\cdot (r^2)^\omega)
=\OOO(r^{2(9+\omega)})$.

We have not yet accounted for the time needed to construct
the graph $G'$ for all $\OOO((r^2)^9)$ choices for the guesses.
Fix $\ell\in [6]$.
The subgraph of $G''$ induced by $\{ z_i^{(\ell)}: i\in [s]\}
\cup \{ z_j^{(\ell+1)}: j\in [s]\}$ can be naively constructed in $\OOO(n(r^2)^2)$ time.  This subgraph is affected by
which vertices in $V^{(\ell+1)}$ are marked ``invalid'', 
which is dependent only on
the guesses for $j[\ell',\ell'']$ with $(\ell',\ell'')\in Z$
and ($\ell'=\ell+1$ or $\ell''=\ell+1$); there are 
$\OOO((r^2)^3)$ possible values for these three indices.
Thus, we can generate these subgraphs over all choices in $\OOO(n(r^2)^2\cdot (r^2)^3) = \OOO(nr^{10})$ time.  From these subgraphs, we
can piece together $G'$ for any sequence of guesses.
\end{itemize}

We set $r=n^{14/(75+8\omega)}$ to equalize $r^{2(9+\omega)}$ with $n^{7/2}/r^{3/4}$.  The term $nr^{10}$ does not dominate.
\end{pproof}

\begin{corollary}\label{cor:segs:Kk}
Let $k$ be a constant integer divisible by~$6$.
Consider a $6$-partite range graph $G=(V,E)$ with $n$ vertices, where each range is a semialgebraic set of constant description complexity in a constant dimension, with intrinsic dimension~$2$.  
We can detect a $k$-clique in $G$ in $\OOO(n^{0.565k})$ time.
\end{corollary}
\begin{pproof}
As in the proof of Corollary~\ref{cor:boxes:Kk},
we apply Theorem~\ref{thm:segs:K6} to a graph with $O(n^{k/6})$
vertices and solve the problem in $\OOO((n^{k/6})^{3.389})$ time.
\end{pproof}

Note that if we were to start from our $\OOO(n^{7/4})$-time
algorithm for $C_3$ in Theorem~\ref{thm:segs:C3}, we would
get a time bound of $O((n^{7k/12})\le O(n^{0.584k})$ for $k$
divisible by 3, which is worse.

Again, there might be room for further small improvement over the above coefficient $0.565$, by analyzing larger patterns beyond $K_6$ with more
effort.

We will see still more algorithms using high-low tricks 
later in Section~\ref{sec:boxes} on independent sets for boxes.

\section{Technique 3: Degeneracy}\label{sec:deg}

In this section, we design faster algorithms exploiting the fact that intersection graphs
avoiding certain patterns are sparse and have low degeneracy.  
Such algorithms are more specialized and do not 
apply to $k$-partite range graphs, however.

Recall that the \emph{degeneracy} of an undirected graph $G$
is defined as the minimum maximum out-degree over
all acyclic orientations of $G$.

\ssubsection{$C_k$ in box intersection graphs for even $k$}

\begin{theorem}\label{thm:boxes:Ck:even}
Let $k\ge 4$ be an even constant.
Given $n$ boxes in a constant dimension,
we can find a $k$-cycle in the intersection graph in $O(T_k(\OO(n),\OO(1)))$ time, where $T_k(m,\Delta)$ denotes
the time complexity for finding a $k$-cycle in a sparse undirected graph with $m$ edges and degeneracy $\Delta$.
\end{theorem}
\begin{pproof}
Let $\D=c\log^{c'} n$ for suitable constants $c,c'$.

We first try to find a subset $V'$ of the vertices whose intersection graph has more than $|V'|\D/2$ edges.
To this end, we generate up to $n\D/2$ edges of the original intersection graph $G=(V,E)$; this takes $\OO(n\D)$ time, by orthogonal range searching.
If the number of edges exceeds $n\D/2$, we can just set $V'=V$ and end the process.
Otherwise, pick a vertex of degree at most $\D$, delete it from the current graph, and repeat.
If we are unable to find such a vertex during an iteration, then we can set $V'$ to be the remaining vertices and end the process.
If we have deleted all vertices, then the original graph $G$ has degeneracy at most $\D$, and so we can solve the
problem in $T_k(n\D/2,\D)$ time.

Having found $V'$, we compute a biclique cover 
$\{(A_1,B_1),\ldots,(A_s,B_s)\}$ of the intersection graph $G'$ of $V'$
of $\OO(|V'|)$ size by Lemma~\ref{lem:bicliques:boxes}.
If $\min\{|A_i|,|B_i|\}\ge k/2$ for some $i$, then we have found a $K_{k/2,k/2}$, which contains a $k$-cycle.
Otherwise, the number of edges in $G'$ is bounded by $\sum_i |A_i||B_i|
\le \OO(\sum_i k(|A_i|+|B_i|))=\OO(|V'|)$, which violates the assumption that the number is more than $|V'|\D/2$, if the constants
$c$ and $c'$ are sufficiently large.
\end{pproof}

According to known results on finding short cycles in low-degeneracy graphs by Alon, Yuster, and Zwick~\cite{AlonYZ97}:
\begin{itemize}
\item $T_4(m,\D) = O(m\D)$ and $T_6(m,\D)=O(m^{3/2}\D^{1/2})$;
\item $T_k(m,\D)=O(m^{2-4/k}\D)$ for $k\equiv 0\bmod 4$, and
$T_k(m,\D)=O(m^{2-4/(k+2)}\D^{1-2/(k+2)})$ for $k\equiv 2\bmod 4$.
\end{itemize}

\ssubsection{$C_4$ in 2D segment intersection graphs}

\begin{theorem}\label{thm:segs:C4}
Given $n$ line segments in $\R^2$, 
we can find a $4$-cycle in the intersection graph
in $\OO(n)$ time. 
\end{theorem}
\begin{pproof}
We follow the same approach as in the proof of Theorem~\ref{thm:boxes:Ck:even} (with $T_4(m,\D) = O(m\D)$).
The main change is that we use 
Lemma~\ref{lem:bicliques:segs} with $k=4$ for the computation of the
$\OO(n)$-size biclique cover.
If the lemma fails to compute a cover but instead finds a $K_k$,
then we have found a 4-cycle.  Also, to generate $\OO(n)$ edges of
the intersection graph in $\OO(n)$ time, we can just use a known output-sensitive algorithm for line-segment intersection~\cite{BerBOOK}.
\end{pproof}

The above argument implies that any intersection graph of line segments
that is $C_4$-free (or more generally $K_{k,k}$-free) must have $\OO(n)$ edges.  This combinatorial result was known before; in fact, the number of edges is $O(n)$ without extra
logarithmic factors~\cite{FoxP08,MustafaP16}, and so the degeneracy is $O(1)$.
It seems plausible that the proof by Mustafa and Pach~\cite{MustafaP16}
could be modified to give 
an alternative algorithm for Theorem~\ref{thm:segs:C4},
without needing biclique covers.

Although our approach also works for other even lengths 
for line-segment intersection graphs, we will give a faster approach in the
next section.

\section{Technique 4: Separators}\label{sec:sep}

In the previous section, we observe that
line-segment intersection graphs avoiding $C_k$ for even $k$
must be sparse.  Sparse intersection graphs in the plane,
like planar graphs, are known to have small separators (e.g., see~\cite{FoxP08} on separators for string graphs).  We exploit this fact to
obtain efficient algorithms for finding $C_k$ by divide-and-conquer. 

\ssubsection{$C_k$ for 2D segment intersection graphs for even $k\ge 6$}

\begin{theorem}\label{thm:segs:Ck:even:sep}
Let $k\ge 6$ be an even constant.
Given $n$ line segments in $\R^2$, 
we can find a $k$-cycle in the intersection graph
in $\OO(n^{\omega/2})\le O(n^{1.187})$ time. 
\end{theorem}
\begin{pproof}
By color coding~\cite{AlonYZ95}, 
it suffices to solve a colored version of the problem,
where the input line segments have been
colored with $k$ colors, and we seek a $k$-cycle under the assumption that 
a $k$-chromatic $k$-cycle exists.  

We solve an extension of the problem: given a set $S$ of $n$ line segments in $\R^2$, and a subset $Q\subset S$ of $q$ line segments, 
(i)~decide whether $S$ has a $k$-chromatic $k$-cycle, and 
(ii)~for every $u,v\in Q$ and every sequence $\gamma$ of length at most $k$, compute the Boolean value $f_S^{(\gamma)}[u,v]$, which is true iff there is a path from $u$ to $v$ in the intersection graph of $S$
whose sequence of colors (excluding the first vertex $u$) is equal to $\gamma$.  The number of different sequences is $O(1)$ (since $k$ is a constant).
The algorithm may stop as soon as a $k$-cycle (not necessarily $k$-chromatic) is found.

First compute a biclique cover 
$\{(A_1,B_1),\ldots,(A_s,B_s)\}$ of the intersection graph $G$ of $S$ of $\OO(n)$ size by Lemma~\ref{lem:bicliques:segs}.
If the lemma fails to compute a cover but instead finds a $K_k$,
then we have found a $k$-cycle and can stop.
If $\min\{|A_i|,|B_i|\}\ge k/2$ for some $i$, then we have found a $K_{k/2,k/2}$, which contains a $k$-cycle, and can stop.
Otherwise, the number of edges in $G$ is bounded by $\sum_i |A_i||B_i|
\le \OO(\sum_i k(|A_i|+|B_i|))=\OO(n)$.  We can afford to explicitly build the graph $G$ (by a segment intersection algorithm in $\OO(n)$ time~\cite{BerBOOK}).

Now, consider the planar graph $H$ formed by the arrangement of $S$,
where the vertices are the $\OO(n)$ segment endpoints and intersection points of $S$.  Initialize the weight of all vertices to 0.
Along each segment $s$ that is incident to $d(s)$ vertices,
add $1/d(s)$ to the weight of each incident vertex.  Then the total
weight is $n$.
Apply the planar separator theorem~\cite{LiptonT80} to partition the vertices of $H$
into $V_1,V_2,V_B$, such that the total weight of $V_i$ is at most $2n/3$ for $i\in\{1,2\}$, and $|V_B|=\OO(\sqrt{n})$, and no pair of
vertices in $V_1\times V_2$ are adjacent; the construction time is $\OO(n)$.

For each $i\in\{1,2\}$,
let $S_i$ be the set of all segments $s\in S$ such that all vertices along
$s$ are in $V_i$.  Let $S_B$ be the set of all segments $s\in S$ such that
$s$ contains a vertex in $V_B$.
Then $|S_1|,|S_2|\le 2n/3$ and $|S_B|=\OO(\sqrt{n})$, and
no pair of segments in $S_1\times S_2$ intersect.

For each $i\in\{1,2\}$, recursively solve the problem for 
$S_i\cup S_B$ with the subset $(Q\cap S_i)\cup S_B$.  After the recursive calls,
we compute $f_S^{(\gamma)}[u,v]$ as follows
(similar to standard approaches to all-pairs shortest paths
or transitive closure by repeatedly squaring matrices):
\begin{enumerate}
\item Initialize $f_S^{(\gamma,0)}[u,v]$ to false for all $u,v,\gamma$.
\item For each $u,v\in (Q\cap S_i)\cup S_B$ and each color sequence $\gamma$ of length at most $k$, if $f_{S_i\cup S_B}^{(\gamma)}[u,v]$ is true,
set $f_S^{(\gamma,0)}[u,v]$ to true.
\item For $j=1,\ldots,\lceil\log k\rceil$ do:
\begin{enumerate}
\item
Initialize $f_S^{(\gamma,j)}[u,v]=f_S^{(\gamma,j-1)}[u,v]$ for all $u,v,\gamma$.
\item
For every $u,v\in Q\cup S_B$ and every pair of color sequences $\gamma$ and
$\gamma'$ with total length at most $k$,
if $\bigvee_{x\in Q\cup S_B} (f_S^{(\gamma,j-1)}[u,x]\wedge f_S^{(\gamma',j-1)}[x,v])$ is true,
then set $f_S^{(\gamma\gamma',j)}[u,v]$ to true,
where $\gamma\gamma'$ denotes the concatenation of $\gamma$ and $\gamma'$.
\end{enumerate}
\end{enumerate}
It is not difficult to see that $f_S^{(\gamma)}[u,v]=f_S^{(\gamma,\lceil\log k\rceil)}[u,v]$ for all $u,v\in Q\cup S_B$.

To detect a $k$-chromatic $k$-cycle in $S$, if one is not found
in the two recursive calls, then it suffices to detect a $k$-chromatic
$k$-cycle that passes through an element of $S_B$.  This can be done by
checking $f_S^{(\gamma)}[u,u]$ for all $u\in S_B$ and all sequences $\gamma$
that are permutations of the $k$ colors.

Each iteration in step~3 can be done by matrix multiplication
in $O(|Q\cup S_B|^\omega) = \OO(q^\omega + n^{\omega/2})$
time.
The total running time satisfies the following recurrence, for
some constant $c$ (for all sufficiently large $n$):
\[ T(n,q)\ = \max_{\scriptsize\begin{array}{c}n_1,n_2\le 0.67n,\ q_1,q_2\le q:\\ n_1+n_2\le n+c\sqrt{n},\\ q_1+q_2\le q+c\sqrt{n}\end{array}} (T(n_1,q_1)+T(n_2,q_2) + 
\OO(q^\omega + n^{\omega/2})).  
\]

Note that if $n_1,n_2\le 0.67n$, $n_1+n_2\le n+c\sqrt{n}$, $q_1,q_2\le q$, and $q_1+q_2\le q+c\sqrt{n}$, it is not difficult to see that
$q_1^2+q_2^2 \le q^2 + (c\sqrt{n})^2$, and so
\begin{eqnarray*}
 (q_1^2 + Cn_1\log n_1) + (q_2^2+Cn_2\log n_2) &\le &
  q^2 + (c\sqrt{n})^2 + C(n+c\sqrt{n})(\log n - \log(1/0.67))\\
  &\le & q^2 + Cn\log n
\end{eqnarray*}
for a sufficiently large constant $C$ (for all sufficiently large $n$).
Thus, $T(n,q)\le T'(n,q^2+Cn\log n)$, where 
\[ T'(n,N)\ = \max_{n_1,n_2\le 0.67n,\ N_1+N_2\le N} (T'(n_1,N_1)+T'(n_2,N_2) + 
\OO(N^{\omega/2})). 
\]
This recurrence solves to $T'(n,N)=\OO(N^{\omega/2})$, and
so $T(n,0)=\OO(n^{\omega/2})$.
\IGNORE{
It is not difficult to see%
\footnote{
Hint: 
$(q_1+C\sqrt{n_1})^\omega + (q_2+C\sqrt{n_2})^\omega
\le (q_1+C\sqrt{0.67n})^\omega + (q_2+C\sqrt{0.67n})^\omega
\le (q+c\sqrt{n}+C\sqrt{0.67n})^\omega + (C\sqrt{0.67n})^\omega
\le (q+C\sqrt{n})^\omega + O(n^{\omega/2})$ for a sufficiently large 
constant $C$.
}
that the recurrence solves to $T(n,q)=\OO((q+\sqrt{n})^{\omega})$, and
so $T(n,0)=\OO(n^{\omega/2})$.
}
\IGNORE{
Let $\Phi=(q+C\sqrt{n})^2$  for a sufficiently large constant $C$.
Let $\Phi_i=(q_i+C\sqrt{n_i})^2$.
Then $\Phi_1+\Phi_2\le (q_1+q_2)^2 + 2C(q_1+q_2)\sqrt{0.67n}+O(n)
\le (q+c\sqrt{n})^2 + 1.8C(q+c\sqrt{n})\sqrt{n} + O(n)
\le \Phi + O(n)$ by choosing $C$ so that $2c+1.8C \le 2C$.
Thus, $T(n,q)\le T'(n,(q+C\sqrt{n})^2)$, where

\[ T'(n,\Phi)\ = \max_{\scriptsize\begin{array}{c}n_1,n_2\le 0.67n:\\ n_1+n_2\le n+c\sqrt{n},\\ \Phi_1+\Phi_2\le \Phi+O(n)\end{array}} (T'(n_1,\Phi_1)+T'(n_2,\Phi_2) + 
\OO(\Phi^{\omega/2})).
\]

The recurrence solves to $T'(n,\Phi)=\OO((\Phi+n)^{\omega/2})$, and
so $T(n,0)=\OO(n^{\omega/2})$.
}
\end{pproof}

Note that the running time above is near linear if $\omega=2$.

\ssubsection{Girth for 2D segment intersection graphs}

The separator-based approach can also be used to compute
the girth (length of the shortest cycle) of line-segment intersection graphs:

\begin{theorem}\label{thm:segs:girth}
Given $n$ line segments in $\R^2$, 
we can compute the girth of the intersection graph
in $\OO(n^{3/2})$ time. 
\end{theorem}
\begin{pproof}
We can check whether the intersection graph $G$ contains a 3-cycle
in $O(n^{1.408})$ time by Theorem~\ref{thm:segs:C3}, or 
a 4-cycle in $\OO(n)$ time by Theorem~\ref{thm:segs:C4}.
So, we may assume that the girth is more than~4.  

By the argument in the proof of Theorem~\ref{thm:segs:Ck:even:sep},
we know that $G$ has $\OO(n)$ edges.  We can afford to explicitly build the graph $G$.

As in the proof of Theorem~\ref{thm:segs:Ck:even:sep},
we can apply the planar separator theorem to
partition the set $S$ of line segments into $S_1,S_2,S_B$,
where $|S_1|,|S_2|\le 2n/3$ and $|S_B|=\OO(\sqrt{n})$, and
no pair of segments in $S_1\times S_2$ intersect.

There are 2 possible cases: (i)~the shortest cycle is entirely 
contained in $S_i$ for some $i\in\{1,2\}$, or (ii)~the shortest cycle
passes through an element of $S_B$.  Case~(i) can be handled
by recursively solving the problem for $S_1$ and for $S_2$.
For case~(ii), we can run breadth-first search from every element $s\in S_B$ in the
intersection graph to find the shortest cycle through $s$.  
Since the graph has $\OO(n)$ edges, the total time for case~(ii) is $\OO(|S_B|n)=
\OO(n^{3/2})$.  The overall running time satisfies the recurrence
\[ T(n)\ = \max_{n_1,n_2\le 2n/3:\ n_1+n_2\le n} (T(n_1)+T(n_2) + 
\OO(n^{3/2})),
\]
which solves to $T(n)=\OO(n^{3/2})$.
\end{pproof}

We leave open the question of whether the running time above could
be further improved to near-linear,
considering that the girth of a planar graph can be computed
in linear time~\cite{ChangL13}.  It is not clear how compute the girth
of a line-segment intersection graph in near-linear time even if it is sparse. In the
arrangement, we are counting
the number of ``turns'' rather than the number of vertices along the cycle,
and certain geometric properties about shortest paths may not hold under
this objective function (for example,
two shortest paths may cross a large number of times, even when they are unique).

\section{Technique 5: Shifted Quadtrees}\label{sec:quadtree}

In this section, we investigate the case of intersection graphs
of fat objects.

There are a number of different ways to define fatness.  We use the
following (from \cite{Chan03}): a family of objects is \emph{fat} if for every hypercube $\gamma$
of side length $r$, the subfamily of all objects intersecting $\gamma$
and having side length at least $r$ can be stabbed by $O(1)$ points.
Here, the \emph{side length} of an object refers to the maximum
side length of its axis-aligned bounding box.

We will adapt a technique by Chan~\cite{Chan03} based on shifted quadtrees and dynamic programming.  This technique was originally used to solve  a 
seemingly different problem: designing
approximation algorithms for independent sets of fat objects.
Interestingly, we show that the technique can be used to find a (non-induced) copy
of any fixed constant-size subgraph (in particular, find $C_k$ or $K_k$) in the intersection graph of fat objects.

\begin{definition}
A \emph{quadtree cell} is a hypercube of the
form $[i_1/2^j,(i_i+1)/2^j)\times\cdots\times [i_d/2^j,(i_d+1)/2^j)$
for integers $i_1,\ldots,i_d,j$.

An object with side length $r$
is \emph{$c$-aligned} if it is contained inside
a quadtree cell of side length at most $cr$.
\end{definition}

The following ``shifting lemma'' is taken from~\cite[Lemma~3.2]{Chan03}
(based on earlier work~\cite{Chan98}):

\begin{lemma}\label{lem:shift}
Let $K>d$ be an odd number.  Let $t^{(j)}=(j/K,\ldots,j/K)\in\R^d$.
For any object $s$ in $[0,1)^d$,
the shifted object $s+t^{(j)}$ is
$(2K)$-aligned for all but at most $d$ indices $j\in [K]$.
\end{lemma}

\ssubsection{Any fixed pattern in fat-object intersection graphs}

\begin{theorem}\label{thm:fat}
Let $k$ be a constant and let $X_k$ be
a graph with $k$ vertices.
Given $n$ fat objects in a constant dimension~$d$, 
we can determine whether $X_k$ is a subgraph of
the intersection graph
in $O(n\log n)$ time. 
\end{theorem}
\begin{pproof}
Let $S$ be the given set of objects.  Assume that
$S\subset [0,1)^d$ (by rescaling).
Assume that the vertices of $X$ are $[k]$ (by relabeling).
We seek an injective mapping $\phi^*: [k]\rightarrow S$
such that for all $i,j\in [k]$, if $ij$ is an edge of $X$,
then $\phi^*(i)$ intersects $\phi^*(j)$.

Let $K=2dk+1$, and let $t^{(j)}=(j/K,\ldots,j/K)\in\R^d$.  By Lemma~\ref{lem:shift} and the pigeonhole principle, there exists $j\in [K]$ such that
$\phi^*(i)+t^{(j)}$ is $(2K)$-aligned for all $i\in[k]$.
We guess such an index~$j$.  (There are only $O(1)$ choices, since
$k$ is a constant.)
Shift all objects by $-t^{(j)}$.  From now on, we can remove
all objects from $S$ that are not $(2K)$-aligned.

\newcommand{\phii}{\widehat{\phi}}
\newcommand{\phiii}{\widetilde{\phi}}

Given a quadtree cell $\gamma$ of side length $r$,
let $S_\gamma$ denote the set of all objects of $S$ contained in $\gamma$,
and let $S_{\partial\gamma}$ denote the set of all objects of $S$
intersecting the boundary of $\gamma$.
Because all objects are $(2K)$-aligned, the objects in $S_{\partial\gamma}$
all have side lengths at least $r/(2K)$.
Because of fatness, we can pierce all these objects with $O(K^{d-1})$ points.  If one of these points pierces more than $k$ objects,
we have found a $K_k$, which contains $X_k$, and can stop.
Thus, we may assume that $|S_{\partial\gamma}|$ is bounded by $O(kK^{d-1})$, which is $O(1)$.   From this property, we can then apply dynamic programming
to solve the problem, intuitively, because the ``interface'' of a quadtree cell has constant size. 

More formally, we define the following collection of subproblems:
\begin{quote}
Given a quadtree cell $\gamma$, 
two subsets of indices
$I\subseteq [k]$ and $I_B\subseteq [k]-I$,
and an injective mapping $\phi: I_B\rightarrow S_{\partial\gamma}$,
we want to compute $f_\gamma[I,I_B,\phi]$, which is true iff
there exists an injective mapping $\phii: I\cup I_B\rightarrow S_\gamma\cup S_{\partial\gamma}$ which is an extension of $\phi$,
with $\phii(I)\subseteq S_\gamma$, such that
for all $i,j\in I\cup I_B$, if $ij$ is an edge of $X$,
then $\phii(i)$ intersects $\phii(j)$.
\end{quote}
Note that there are only $O(1)$ choices of $I$, $I_B$, and $\phi$.
The original problem corresponds to the case when $\gamma=[0,1)^d$,
$I=[k]$, and $I_B=\emptyset$.

To compute $f_\gamma[I,I_B,\phi]$:  If $S_\gamma$ has constant size,
the computation trivially takes $O(1)$ time.  Otherwise, ``shrink''
$\gamma$ to the smallest
quadtree cell $\gamma'$ containing all of $S_\gamma$, and then
split $\gamma'$ into $D := 2^d$ quadtree cells $\gamma_1,\ldots,\gamma_{D}$.
Initialize $f_\gamma[I,I_B,\phi]$ to false.
Examine each possible partition of $I$ into subsets
$I_1,\ldots,I_D,I_B'$ such that no pairs of indices in $I_i\times I_j$
for distinct $i,j\in [D]$ are adjacent in $X$.
Examine each possible injective mapping $\phiii: I_B'\cup I_B\rightarrow 
(S_\gamma \cap (S_{\partial\gamma_1}\cup\cdots\cup S_{\partial\gamma_{D}}))\cup S_{\partial\gamma}$ which is an extension of $\phi$,
with $\phiii(I_B')\subseteq S_\gamma\cap (S_{\partial\gamma_1}\cup\cdots\cup S_{\partial\gamma_{D}})$.
There are $O(1)$ choices for $I_1,\ldots,I_D,I_B',\phiii$.
Let $(I_i)_B = \{i\in I_B'\cup I_B: \phiii(i)\in S_{\partial\gamma_i}\}$.
Let $\phiii_i$ be the restriction of $\phiii$ to $I_i\cup (I_i)_B$.
If $\bigwedge_{i=1}^D f_{\gamma_i}[I_i,(I_i)_B, \phiii_i]$ is true for some choice,
then set $f_\gamma[I,I_B,\phi]$ to true.

To analyze the running time, let $P$ be the set of corner points of the
objects' bounding boxes. First note that the quadtree cells
generated by the above recursion correspond precisely to the nodes
of the \emph{compressed quadtree}~\cite{Har11} of $P$ (degree-1 nodes are eliminated because
we ``shrink'' the cell before splitting).  It is known that
the compressed quadtree can be constructed in $O(n\log n)$ time
(in fact, $O(n)$ time~\cite{Chan08} if the coordinate values of $P$ are $O(\log n)$ bits long), assuming a reasonable computational model.
There are $O(n)$ cells $\gamma$ in the tree.
At each quadtree cell~$\gamma$, we need to explicitly generate
$S_{\partial\gamma}$.  To this end, for each object $s\in S$,
we find the lowest common ancestor (LCA) of the corner points of $s$ in the
quadtree (the tree may not be balanced, but LCAs can still
be done in $O(1)$ time~\cite{BenderF00}).  
We can then descend from the LCA to find all nodes in the tree whose 
cells intersect the boundary of $s$, in time proportional to the output size.  Since the total size of $S_{\partial\gamma}$ is $O(n)$, the total time for this step is $O(n)$.
Afterwards, the above procedure allows us to compute all $f_\gamma$ values of
a cell $\gamma$ from the $f_{\gamma_i}$ values of its children cells $\gamma_i$ in $O(1)$ time per cell.  By evaluating bottom-up, the 
computation takes $O(n)$ time.
\end{pproof}



\section{Technique 6: Round-Robin Recursion}\label{sec:recurs}

In this section, we propose yet another technique for finding
$C_3$ or $I_3$.  It is based on a proof by Agarwal and Sharir~\cite{AgarwalS02} (see also \cite{FranklK21}) for 
a combinatorial problem (bounding the number of congruent copies
of a fixed simplex in a point set).  The technique uses
a round-robin recursion in combination with cuttings.

The result is limited to a special class of range graphs,
where the actual dimension (not the intrinsic dimension) of the ranges is small, namely, 2 or 3.

\ssubsection{$C_3$ (or $I_3$) in 2D/3D translates intersection graphs}

\begin{theorem}\label{thm:translates}
Consider a tripartite graph $G=(V,E)$ with $n$ vertices, where $V$ is divided into parts $V^{(1)},V^{(2)},V^{(3)}$.
Each point $v\in V$ is associated with a point $p(v)\in\R^d$,
and semialgebraic sets $R^{(\ell\ell')}(v)\subseteq \R^d$ of constant description complexity for each $\ell,\ell\in[3]$, such that
for each $u\in V^{(\ell)}$ and $v\in V^{(\ell')}$ with $\ell\neq\ell'$,
we have $uv\in E$ iff $p(u)\in R^{(\ell\ell')}(v)$.

Then we can find a $3$-cycle in $G$ in $\OOO(n^{3/2})$ time if $d=2$,
or $\OOO(n^{9/5})$ time if $d=3$.
\end{theorem}
\begin{pproof}
We seek a 3-cycle $v_1^*v_2^*v_3^*$ with $v_1^*\in V^{(1)},\ldots,
v_3^*\in V^{(3)}$.
Let $n_\ell=|V^{(\ell)}|$ for $\ell\in[3]$.
Let $r$ be a parameter.  

Let $S_2=\{R^{(12)}(v_2): v_2\in V^{(2)}\}$ and $S_3=\{R^{(13)}(v_3):
 v_3\in V^{(3)}\}$.
By standard geometric sampling techniques~\cite{Clarkson87,ChazelleF90}, for $d\le 3$, we can decompose space into
$r^{d+o(1)}$ cells (a \emph{cutting}), such that each cell is intersected by the boundaries of at most $n_2/r$ 
ranges of $S_2$ and  at most $n_3/r$ 
ranges of $S_3$.  The construction takes $O(r^{O(1)}n)$ time.
We can further subdivide each cell into subcells (by extra vertical cuts) so that
each subcell contains at most $n_1/r^d$ points in $\{p(v_1): v_1\in V^{(1)}\}$; the number of cells remains $r^{d+o(1)}$.

For each cell $\gamma$, let $V_\gamma^{(1)}=\{v_1\in V^{(1)}: p(v_1)\in\gamma\}$, and for $\ell\in\{2,3\}$, let
$V_\gamma^{(\ell)}=\{v_\ell\in V^{(\ell)}: \mbox{the boundary of $R^{(1\ell)}(v_\ell)$ intersects $\gamma$}\}$ and
let $Y_\gamma^{(\ell)}=\{v_\ell\in V^{(\ell)}: \mbox{$R^{(1\ell)}(v_\ell)$ contains $\gamma$}\}$. 

We guess the cell $\gamma$ with $v_1^*\in V_\gamma^{(1)}$.  We consider 
three cases:
\begin{itemize}
\item {\sc Case 1:} $v_2^*\in V_\gamma^{(2)}$, $v_3^*\in V_\gamma^{(3)}$.
We just recursively solve the problem for $(V_\gamma^{(1)},V_\gamma^{(2)},V_\gamma^{(3)})$.
\item {\sc Case 2:} $v_2^*\in Y_\gamma^{(2)}$.
We know there is an edge between every $v_1\in V_\gamma^{(1)}$
and every $v_2\in Y_\gamma^{(2)}$.
For each $v_3\in V^{(3)}$, 
we test whether there exists $v_1\in V_\gamma^{(1)}$
such that $v_1v_3\in E$, i.e., $p(v_1)$ is in the range $R^{(13)}(v_3)$,
and similarly test whether there exists $v_2\in Y_\gamma^{(2)}$
such that $v_2v_3\in E$, i.e., $p(v_2)$ is in the range $R^{(23)}(v_3)$.
If both tests return true, we have found a 3-cycle $v_1v_2v_3$.
These tests reduce to $O(n_3)$ range queries on $O(n_1+n_2)$ points in $d$ dimensions.
By known range searching results~\cite{AgaEriSURV,Matousek94survey,Agarwal17survey}, the time complexity is at most
$\OOO((n_1+n_2+n_3)^{2d/(d+1)})$.

\item {\sc Case 3:} $v_3^*\in Y_\gamma^{(3)}$.
Similar to Case 2.
%
\end{itemize}

The total running time satisfies the following recurrence:
\begin{eqnarray*}
 T(n_1,n_2,n_3) &\le& r^{d+o(1)} T(n_1/r^d, n_2/r, n_3/r) + \OOO(r^{O(1)} (n_1+n_2+n_3)^{2d/(d+1)}).
\end{eqnarray*}
By symmetry, we have the following similar recurrences: 
\begin{eqnarray*}
T(n_1,n_2,n_3) &\le& r^{d+o(1)} T(n_1/r, n_2/r^d, n_3/r) + \OOO(r^{O(1)} (n_1+n_2+n_3)^{2d/(d+1)})\\ 
T(n_1,n_2,n_3) &\le& r^{d+o(1)} T(n_1/r, n_2/r, n_3/r^d) + \OOO(r^{O(1)} (n_1+n_2+n_3)^{2d/(d+1)}).
\end{eqnarray*}
Applying the three in a round-robin manner, we get the recurrence
\[ T(n,n,n)\ \le\ r^{3d+o(1)} T(n/r^{d+2},n/r^{d+2},n/r^{d+2}) + \OOO(r^{O(1)} n^{2d/(d+1)}),
\]
which solves to $T(n,n,n) = O(n^{3d/(d+2)+\eps})$ by making $r$ an arbitrarily large constant.
\end{pproof}

Besides the assumption the actual dimension $d$ is small
instead of the intrinsic dimension,
another subtle difference in the above theorem is that
each vertex $v$ is associated
with just one point $p(v)$, not multiple points $p^{(\ell\ell')}(v)$.
These requirements limit the applicability of the theorem.
On the other hand, the result is still interesting, considering
the lack of any
subquadratic results for nonorthogonal, nonfat objects in 3D before this section.

One class of graphs that satisfy these requirements is
intersection graphs of translates of a fixed shape $S_0$ in 2D and 3D\@.  
The shape need not be
fat or convex; it just needs to be semialgebraic with constant description
complexity.  We simply note that $S_0+u$ intersects $S_0+v$ iff
the point $u$ lies in the region $(S_0+v)-S_0$,
or equivalently, the point $v$ lies in the region $(S_0+u)-S_0$.
(For independent set, we take the complement of the region.)

For another application, we can solve the following problem:
given a set $P$ of $n$ points in $\R^d$ with $d\in\{2,3\}$ and fixed
constants $a,b,c,\delta$, find
3 points $p_1,p_2,p_3\in P$ such that
$\|p_1-p_2\|\in (a-\delta,a+\delta)$,
$\|p_2-p_3\|\in (b-\delta,b+\delta)$,
and $\|p_3-p_1\|\in (c-\delta,c+\delta)$.
Here, the ranges of interest are fixed-radii annuli or spherical shells.
This problem can be viewed as an approximate geometric pattern matching
problem: finding three points that almost match a fixed triangle $T_0$
with side lengths $a,b,c$, allowing translations and rotations.  (A more standard version of approximate geometric pattern matching measures error in terms of Hausdorff distance~\cite{GoodrichMO99},
but it can be shown that the two versions are related, if
$T_0$ is fat, such as an equilateral triangle.
For $d=2$, the known algorithm for the Hausdorff-distance version has $\OO(n^2)$ running time,
but our algorithm has $\OO(n^{3/2})$ running time.)

\section{Independent Sets for Boxes}\label{sec:boxes}

We now focus on finding small-size independent sets for boxes,
aiming for faster algorithms than the more general
algorithms from Sections \ref{sec:bicliques}--\ref{sec:high:low}.

The following trivial fact will be useful: two boxes are independent iff they
are separable by an axis-aligned hyperplane.  Our results will 
be obtained by using
orthogonal range searching data structures and applying
more high-low tricks.  Biclique covers are no longer needed.

\ssubsection{$I_3$ in box intersection graphs}

\begin{theorem}\label{thm:boxes:I3}
Given $n$ red/blue/green boxes in a constant dimension~$d$, we can find a $3$-chromatic size-$3$ independent set in $\OO(n)$ time. 
\end{theorem}
\begin{pproof}
Let $V^{(1)},V^{(2)},V^{(3)}$ be the color classes.
We seek an independent set $\{v_1^*,v_2^*,v_3^*\}$ with $v_1^*\in V^{(1)}, \ldots, v_3^*\in V^{(3)}$.
We know that the boxes $v_2^*$ and $v_3^*$ must be separated by an
axis-aligned hyperplane; say it is orthogonal to the $x$-th coordinate axis, and that
the right $x$-th coordinate of $v_2^*$
is smaller than the left $x$-th coordinate of $v_3^*$.
(The other cases are similar; we can try them all.)

For each $v_1\in V^{(1)}$,
we find a box $v_2\in V^{(2)}$ that does not intersect $v_1$,
while minimizing its right $x$-th coordinate.
This can be found by an orthogonal range
minimum query~\cite{AgaEriSURV} in a constant dimension.
Similarly, we find a box $v_3\in V^{(3)}$ that does not intersect $v_1$,
while maximizing its left $x$-th coordinate.
If these two boxes $v_2$ and $v_3$ do not intersect, we have found a solution $\{v_1,v_2,v_3\}$.  
(To summarize: once $v_1$ is fixed, we can greedily generate
candidates for $v_2$ and $v_3$.)
In total, we have made $O(n)$ range queries on $O(n)$ elements,
requiring $\OO(n)$ time.
\end{pproof}

\ssubsection{$I_4$ in box intersection graphs}

\begin{theorem}\label{thm:boxes:I4}
Given $n$ boxes in a constant dimension~$d$, colored with $4$ colors, we can find a
$4$-chromatic size-$4$ independent set in $\OO(n^{3/2})$ time. 
\end{theorem}
\begin{pproof}
Let $G=(V,E)$ denote the complement of the intersection graph.
Let $V^{(1)},\ldots,V^{(4)}$ be the color classes.
We seek a 4-clique $\{v_1^*,\ldots,v_4^*\}$ in $G$ with $v_1^*\in V^{(1)}, \ldots, v_4^*\in V^{(4)}$.
For each $\ell,\ell'\in [4]$ with $\ell<\ell'$,
we know that the boxes $v_\ell^*$ and $v_{\ell'}^*$ must be separated by an
axis-aligned hyperplane; say it is orthogonal to the $\xi_{\ell\ell'}$-th coordinate axis.  We guess each $\xi_{\ell\ell'}\in [d]$.
(There are only $O(1)$ guesses; we can try them all.)
Also assume that the right $\xi_{\ell\ell'}$-th coordinate of $v_\ell^*$
is smaller than the left $\xi_{\ell\ell'}$-th coordinate of $v_{\ell'}^*$.
(Again, we can try all cases.)

Let $r\le n$ be a parameter.
Divide the real line into $r$ intervals, each containing
$O(n/r)$ of the $\xi_{12}$-th coordinates of the boxes.
Call an edge $v_1v_2\in E$ \emph{low} if the right $\xi_{12}$-th coordinate of $v_1$ and the left $\xi_{12}$-th coordinates of $v_2$ are in the same interval,
and \emph{high} otherwise.  The number of low edges is $O(n^2/r)$.

\begin{itemize}
\item {\sc Case 1}: $v_1^*v_2^*$ is low.
Fix a low edge $v_1v_2$ with $v_1\in V^{(1)}$ and $v_2\in V^{(2)}$.
We find a box $v_3\in V^{(3)}$ that does not intersect boxes $v_1$ and $v_2$, 
while minimizing its right $\xi_{34}$-th coordinate.
This can be found by an orthogonal range
minimum query in a constant dimension.
Similarly, we find a box $v_4\in V^{(4)}$ that does not intersect boxes $v_1$ and $v_2$, 
while maximizing its left $\xi_{34}$-th coordinate.
If these two boxes $v_3$ and $v_4$ do not intersect, we have found a solution $\{v_1,\ldots,v_4\}$.  
Totalling over all low edges $v_1v_2$, we have made $O(n^2/r)$ range queries on $O(n)$ elements,
requiring $\OO(n^2/r)$ time.

\item {\sc Case 2}: $v_1^*v_2^*$ is high.
Then the $\xi_{12}$-th coordinates
of $v_1^*$ and $v_2^*$ are separated by one of the endpoints of the $r$ intervals.
We guess this endpoint $x$.  There are $O(r)$ choices for $x$.
Fix $v_4\in V^{(4)}$.
\begin{itemize}
\item We find a box $v_1\in V^{(1)}$ that has right $\xi_{12}$-th coordinate less than $x$ and does not intersect $v_4$,
while minimizing its right $\xi_{13}$-th coordinate.
This can be found by an orthogonal range
minimum query.
\item Similarly, 
we find a box $v_2\in V^{(2)}$ that has right $\xi_{12}$-th coordinate greater than $x$ and does not intersect $v_4$,
while minimizing its right $\xi_{23}$-th coordinate.
\item Then we attempt to find a box $v_3\in V^{(3)}$ that does not intersect
$v_4$, and has left $\xi_{13}$-th coordinate greater than
$v_1$'s right $\xi_{13}$-th coordinate, and has
left $\xi_{23}$-th coordinate greater than 
$v_2$'s right $\xi_{23}$-th coordinate.
\end{itemize}
If such a $v_3$ exists, we have found a solution $\{v_1,\ldots,v_4\}$, since none of $v_1,v_2,v_3$ intersects with $v_4$, and $v_1$ and $v_2$ 
don't intersect, and $v_1$ and $v_3$ don't intersect, and $v_2$ and $v_3$ don't intersect.
(To summarize: intuitively, once $v_4$ is fixed, we seek a 3-cycle $v_1v_2v_3$;
and once a separating hyperplane between
$v_1$ and $v_2$ is fixed, we just seek a path $v_1v_3v_2$,
and we can greedily generate candidates for $v_1$ and $v_2$.)
Totaling over all $v_4\in V^{(4)}$, we have made $O(n)$ range queries
on $O(n)$ elements, requiring $\OO(n)$ time.
Summing over all $O(r)$ guesses gives a time bound of $\OO(nr)$.
\end{itemize}

We set $r=\sqrt{n}$ to equalize $nr$ with $n^2/r$.
\end{pproof}

\ssubsection{$I_4$ in 5D box intersection graphs}

\begin{theorem}\label{thm:boxes:I4:5d}
Given $n$ boxes in $\R^5$, colored with $4$ colors, we can find a
$4$-chromatic size-$4$ independent set in $\OO(n)$ time. 
\end{theorem}
\begin{pproof}
Let $G=(V,E)$ denote the complement of the intersection graph.
Let $V^{(1)},\ldots,V^{(4)}$ be the color classes.
We seek a 4-clique $\{v_1^*,\ldots,v_4^*\}$ in $G$ with $v_1^*\in V^{(1)}, \ldots, v_4^*\in V^{(4)}$.
For each $\ell,\ell'\in [4]$ with $\ell<\ell'$,
we know that the boxes $v_\ell^*$ and $v_{\ell'}^*$ must be separated by an
axis-aligned hyperplane; say it is orthogonal to the $\xi_{\ell\ell'}$-th coordinate axis.
We guess each $\xi_{\ell\ell'}\in [5]$.
(There are only $O(1)$ guesses; we can try them all.)
Also guess that the right $\xi_{\ell\ell'}$-th coordinate of $v_\ell^*$
is smaller than the left $\xi_{\ell\ell'}$-th coordinate of $v_{\ell'}^*$.
(Again, we can try all cases.)

By the pigeonhole principle, among the six elements $\xi_{\ell\ell'}\in [5]$, two of them must be equal, say, to 1.  When projecting $v_1^*,v_2^*,v_3^*,v_4^*$ to the 1st coordinate axis, we see two pairs of
disjoint intervals (the two pairs may or may not share an interval);
for example, see Figure~\ref{fig:intervals}. 
It follows that (i)~one interval is completely to the left of two other
intervals, or (ii)~one interval is completely to the right of two other
intervals.  W.l.o.g., say (i) is true.
Suppose that $v_1^*$'s right 1st coordinate is less than
$v_4^*$'s left 1st coordinate, which in turn is less than $v_2^*$'s left 1st coordinate.
(All other cases are similar; we can try them all.)

\begin{figure}[h]
\centering
\includegraphics[scale=0.8]{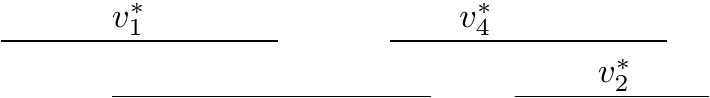}
\caption{Two pairs of disjoint intervals.}\label{fig:intervals}
\end{figure}

The idea is that if we ensure that $v_1$'s right 1st coordinate
is less than $v_4$'s left 1st coordinate, and $v_4$'s left 1st
coordinate is less than $v_2$'s left 1st coordinate, then
it is unnecessary to enforce the condition that $v_1$ and $v_2$
do not intersect.

Fix $v_4\in V^{(4)}$. 
\begin{itemize}
\item[]
\begin{itemize}
\item
We find a box $v_1\in V^{(1)}$ that has right 1st coordinate less than 
$v_4$'s left 1st coordinate,
while minimizing its right $\xi_{13}$-th coordinate.
This can be found by an orthogonal range
minimum query.
\item
Similarly, 
we find a box $v_2\in V^{(2)}$ that has left 1st coordinate greater than $v_4$'s left 1st coordinate,
while minimizing its right $\xi_{23}$-th coordinate.
\item
Then we attempt to find a box $v_3\in V^{(3)}$ that does not intersect
$v_4$, and has left $\xi_{13}$-th coordinate greater than
$v_1$'s right $\xi_{13}$-th coordinate, and has
left $\xi_{23}$-th coordinate greater than 
$v$'s right $\xi_{23}$-th coordinate.
\end{itemize}
\end{itemize}
If such a $v_3$ exists, we have found a solution $\{v_1,\ldots,v_4\}$.
(To summarize: intuitively, once $v_4$ is fixed, we seek a 3-cycle $v_1v_2v_3$; and
since the disjointness between $v_1$ and $v_2$ is automatically
enforced, we just seek a path $v_1v_3v_2$, and so again we can greedily
guess candidates for $v_1$ and $v_2$.)
Totaling over all $v_4\in V^{(4)}$, we have made $O(n)$ range queries
on $O(n)$ elements, requiring $\OO(n)$ time.
\end{pproof}

\ssubsection{$I_5$ in 2D box intersection graphs}

For an axis-aligned rectangle $s=[x_1,x_2]\times [y_1,y_2]$,
define its \emph{SW (resp.\ NE) extension} to be the quadrant $(-\infty,x_2]\times(-\infty,y_2]$
(resp.\ $[x_1,\infty)\times[y_1,\infty)$).

\begin{lemma}\label{lem:I}
Given a set $I$ of disjoint axis-aligned rectangles in $\R^2$,
there exists a rectangle $s\in I$ whose SW extension is 
disjoint from all other rectangles in $I$.
\end{lemma}
\begin{pproof}
Call a rectangle in $I$ \emph{good} if its bottom side is completely visible from below.  For example, the rectangle with the lowest
bottom side is good.  Let $s$ be the good rectangle with the leftmost
left side.  If the SW extension of $s$ intersects another rectangle of $I$, then let $s'$ be such a rectangle having the lowest bottom side.
This rectangle $s'$ is good and is to the left of $s$,
contradicting the leftmost choice of $s$.
\end{pproof}

\begin{theorem}\label{thm:boxes:I5:2d}
Given $n$ axis-aligned rectangles in $\R^2$ colored with $5$ colors, we can find a
$5$-chromatic size-$5$ independent set in $\OO(n^{4/3})$ time. 
\end{theorem}
\begin{pproof}
Let $G=(V,E)$ denote the complement of the intersection graph.
Let $V^{(1)},\ldots,V^{(5)}$ be the color classes.
We seek a 5-clique $K^*=\{v_1^*,\ldots,v_5^*\}$ in $G$ with $v_1^*\in V^{(1)}, \ldots, v_5^*\in V^{(5)}$.

Let $r\le n$ be a parameter.
Form an $r\times r$ grid, where
each column contains $O(n/r)$ vertical sides of the rectangles and each row contains
$O(n/r)$ horizontal sides (by standard perturbation arguments, we may assume that all coordinate values are distinct).
Call an edge $v_1v_2\in E$ \emph{low} if one of the vertical sides
of $v_1$ and one of the vertical sides of $v_2$ lie in the same column,
or one of the horizontal sides
of $v_1$ and one of the horizontal sides of $v_2$ lie in the same row.
The number of low edges is $O(n^2/r)$.

\begin{itemize}
\item {\sc Case 1}: At least one edge of $K^*$ is low.
Say it is $v_1^*v_2^*$.
Among the 3 disjoint rectangles $v_3^*,v_4^*,v_5^*$, there is a
vertical or horizontal line separating one from the other two (because we are in 2D).
W.l.o.g, assume that the separating line is vertical,
with $v_3^*$ to the left, and $v_4^*$ and $v_5^*$ to the right,
and with $v_4^*$ below $v_5^*$.
(All other cases can be handled similarly.)

Fix a low edge $v_1v_2$ with $v_1\in V^{(1)}$ and $v_2\in V^{(2)}$.
We find a rectangle $v_3\in V^{(3)}$ that does not intersect $v_1$ and $v_2$, having the leftmost right side.
This can be found by an orthogonal range
minimum query in a constant dimension.
Next, we find a rectangle $v_4\in V^{(4)}$ that does not intersect $v_1$ and $v_2$, and is to right of the right side of $v_3$, with
the lowest top side.
Similarly, we find a rectangle $v_5\in V^{(5)}$ that does not intersect  $v_1$ and $v_2$, and is to right of the right side of $v_3$, with
the highest bottom side.
These can also be found by orthogonal range searching.
If these two rectangles $v_4$ and $v_5$ do not intersect, we have found a solution $\{v_1,\ldots,v_5\}$.  
Totaling over all low edges $v_1v_2$, we have made $O(n^2/r)$ range queries on $O(n)$ elements,
requiring $\OO(n^2/r)$ time.

\item {\sc Case 2}: All edges of $K^*$ are high.
By Lemma~\ref{lem:I}, we can take the SW extension of one of the
rectangles in $K^*$---say it is $v_1^*$---without intersecting the other rectangles of $K^*$.  Similarly,  we can take the NE extension of one of the
rectangles in $K^*$---say it is $v_2^*$---without intersecting the other rectangles of $K^*$.  (It is easy to see that the these two rectangles can't be the same.)  For example, see Figure~\ref{fig:rects}(a).

\begin{figure}
\centering
\includegraphics[scale=0.6]{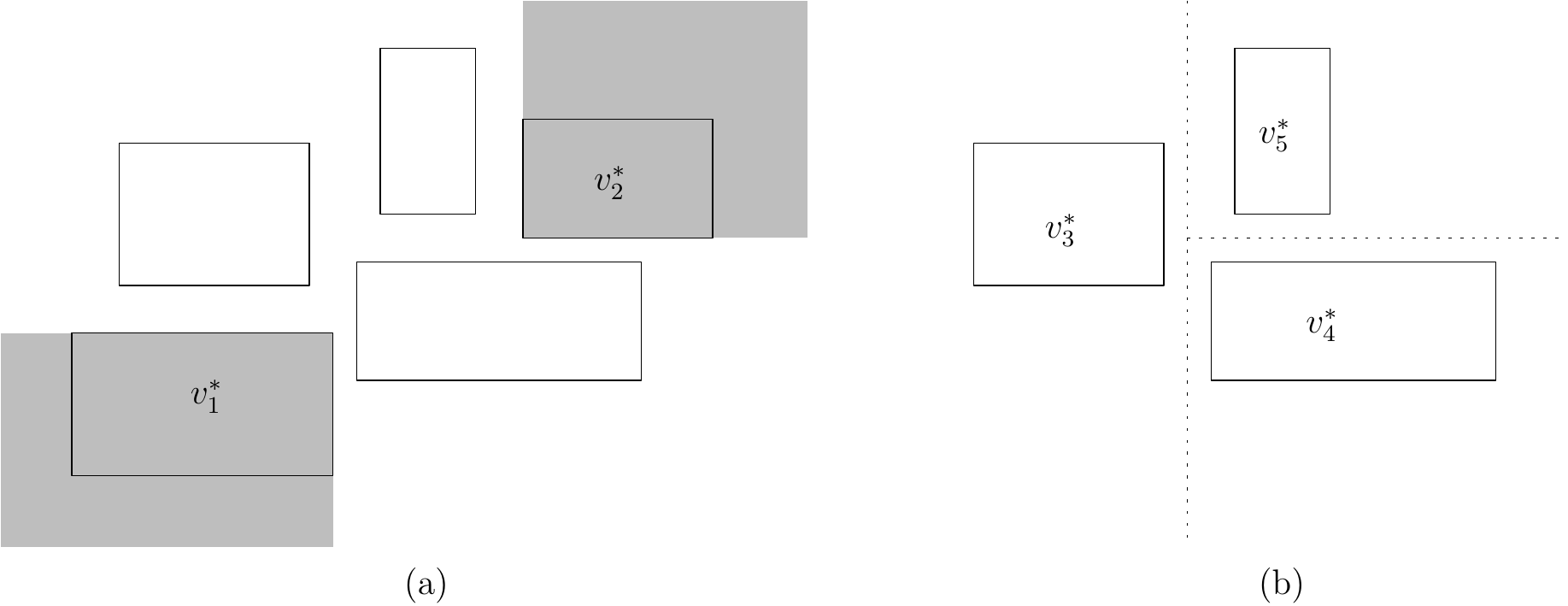}
\caption{(a)~A size-5 independent set of rectangles in $\R^2$.  (b)~A size-3 independent set.}\label{fig:rects}
\end{figure}

A SW quadrant $(-\infty,x]\times (-\infty,y]$ 
(resp.\ NE quadrant $[x,\infty)\times [y,\infty)$) is called \emph{grid-aligned}
if $(x,y)$ is one of the $r^2$ grid points.
Let $\rho_1^*$ be the smallest grid-aligned SW quadrant containing 
the SW extension of $v_1^*$ (in other words, we ``round'' $v_1^*$'s corner point  upward and rightward to
a grid point),
and let $\rho_2^*$ be the smallest grid-aligned NE quadrant containing 
the NE extension of $v_2^*$  (in other words, we ``round'' $v_2^*$'s corner point downward and leftward to
a grid point). 
The rectangles $\rho_1^*,\rho_2^*,v_3^*,v_4^*,v_5^*$ are pairwise disjoint (because all edges of $K^*$ are high).
Among $v_3^*,v_4^*,v_5^*$, there is a
vertical or horizontal line separating one from the other two.
W.l.o.g, assume that the separating line is vertical,
with $v_3^*$ to the left, and $v_4^*$ and $v_5^*$ to the right,
and with $v_4^*$ below $v_5^*$.  For example, see Figure~\ref{fig:rects}(b).

Fix a grid-aligned SW quadrant $\rho_1$ that contains at least one rectangle $v_1\in V^{(1)}$.
Fix a grid-aligned NE quadrant $\rho_2$ that contains at least one rectangle $v_2\in V^{(2)}$, and does not intersect $\rho_1$.
We find a rectangle $v_3\in V^{(3)}$ that does not intersect $\rho_1$ and $\rho_2$, with the leftmost right side.
This can be found by an orthogonal range
minimum query in a constant dimension.
Next, we find a rectangle $v_4\in V^{(4)}$ that does not intersect $\rho_1$ and $\rho_2$, and is to right of the right side of $v_3$, with
the lowest top side.
Similarly, we find a rectangle $v_5\in V^{(5)}$ that does not intersect $\rho_1$ and $\rho_2$, and is to right of the right side of $v_3$, with
the highest bottom side.
These can also be found by orthogonal range searching.
If these two rectangles $v_4$ and $v_5$ do not intersect, we have found a solution $\{v_1,\ldots,v_5\}$.  

There are $O(r^2)$ choices of $\rho_1$.  However, it suffices to consider
the \emph{minimal} grid-aligned SW quadrants $\rho_1$ that contain at least one rectangle $v_1\in V^{(1)}$, and there are only $O(r)$ choices of such $\rho_1$, since the corners of the minimal quadrants form a staircase in the grid.  Similarly, it suffices to consider $O(r)$ choices of $\rho_2$.
Totaling over all $O(r)$ choices of $\rho_1$ and
$O(r)$ choices of $\rho_2$, we have made $O(r^2)$ range queries on $O(n)$ elements,
requiring $\OO(n+r^2)$ time.
\end{itemize}

We set $r=n^{2/3}$ to equalize $r^2$ with $n^2/r$.
\end{pproof}

\section{Final Remarks}

There are still many variants of these
problems that we could address but have not in this paper.
We briefly mention the following:
\begin{itemize}
\item \emph{Counting.} 
For example, the reduction in Theorem~\ref{thm:boxes:Ck} 
still works for counting instead of detection.  And the algorithm in Theorem~\ref{thm:segs:C3}
also seems suitable for counting.
\item \emph{Minimizing weight.}
For \emph{vertex} weights, the reduction in Theorem~\ref{thm:boxes:Ck} 
still works, and we can use known results on minimum-weight 3-cycles for sparse graphs~\cite{CzumajL09}.  And the algorithm in Theorem~\ref{thm:segs:C3}, also seems suitable
by using range minimum queries.  
Kaplan et al.~\cite{KaplanKMRSS19} considered finding the minimum-weight $C_3$ in
a disk intersection graph
with \emph{edge} weights that are Euclidean distances between the centers;
our algorithm in Theorem~\ref{thm:fat} also seems adaptable to such
edge weights, with some modifications.
\item \emph{Other geometric objects.}
Lemmas~\ref{lem:bicliques:segs} and~\ref{lem:segs:I2}
hold not only for line segments but also for triangles or constant-size
convex polygons (with more polylogarithmic factors).  So, the algorithm
in Theorem~\ref{thm:segs:I3} can be generalized to these objects.
(One could also consider certain families of curves: Lemma~\ref{lem:bicliques:segs} holds for pseudo-line segments, and good separators still exist for sparse intersection graphs of curves, but not all techniques for line segments are applicable without increasing the time bounds.)
\item \emph{Other combinations of patterns and objects.}
For example, we have not yet considered
$I_4$ for line-segment intersection graphs,
$K_5$ for boxes or line segments, $K_4$ for 2D translates, etc.
\end{itemize}

For many of the specific results obtained in this paper, there is room
for further improvement in the upper bounds.  It would also be desirable to prove
conditional lower bounds for more problems in this class.

A main open question is whether subquadratic algorithms exist
for $C_3$ in range graphs with intrinsic dimension~3 and higher.
Our algorithms work in higher dimensions, but the time bound
becomes superquadratic (for intrinsic dimension~3, we might still be able to outperform
known general $O(n^\omega)$-time algorithms under the current
value of $\omega$, but the time bound would certainly 
be worse for intrinsic dimension~4).  Similarly, the
algorithm in Theorem~\ref{thm:translates} becomes superquadratic
for dimension
4 and higher.

For one specific open question: can length-3 \emph{depth cycles}~\cite{ChazelleEGPSSS91} be detected
in subquadratic time for a given set of $n$ lines in $\R^3$?
This problem reduces to finding $C_3$ in a range graph with dimension 4.

\appendix

\section{Appendix}

\ssubsection{Proof of Lemma~\ref{lem:bicliques:2d:0}}\label{app:bicliques:2d:0}

The proof is based on multi-level cutting trees~\cite{Clarkson87,AgaEriSURV,Matousek94survey}.

Assume that $V$ is divided into two parts $V^{(1)}$ and $V^{(2)}$.
Let $n_1=|V^{(1)}|$ and $n_2=|V^{(2)}|$.
Assume that each range $R^{(12)}(v)\subseteq \R^d$ is expressed as a
fixed Boolean combination of $\ell$ polynomial inequalities
$f_v(x_{\xi_1},x_{\eta_1})\ge 0$, \ldots, $f_v(x_{\xi_\ell},x_{\eta_\ell})\ge 0$
(of constant description complexity) over the variables $(x_1,\ldots,x_d)$,
for a constant $\ell$ and a fixed sequence of indices $\xi_1,\eta_1,\ldots,
\xi_\ell,\eta_\ell\in [d]$.  There are only $O(1)$ number of such
index sequences, and $O(1)$ number of Boolean combinations.

Let $b$ be a parameter.
By standard results on \emph{cuttings}~\cite{Clarkson87,ChazelleF90,AgaEriSURV}, we can divide the plane into
$b^{2+o(1)}$ cells, such that each cell crosses the boundaries of at most
$n_2/b$ regions $\{(x_{\xi_\ell},x_{\eta_\ell})\in\R^2: f_v(x_{\xi_\ell},x_{\eta_\ell})\ge 0\}$
over all $v\in V^{(2)}$.
The construction takes $O(b^{O(1)}n_2)$ time.
We can further subdivide each cell (by extra vertical cuts) so that
each subcell contains at most $n_1/b^2$ points in $\{(x_{\xi_\ell},x_{\eta_\ell}):
p^{(12)}(u)=(x_1,\ldots,x_d),\ u\in V^{(1)}\}$;
the number of cells remains $b^{2+o(1)}$.

For each cell $\gamma$, let  $V^{(1)}_\gamma = \{u\in V^{(1)}:
p^{(12)}(u)=(x_1,\ldots,x_d),\ (x_{\xi_\ell},x_{\eta_\ell})\in\gamma\}$,
let $V^{(2)}_\gamma = \{u\in V^{(2)}:\ \mbox{$f_v=0$ intersects $\gamma$}\}$,
let $Y^{(2)}_\gamma = \{u\in V^{(2)}:\ \mbox{$f_v > 0$ contains $\gamma$}\}$, and
let $Z^{(2)}_\gamma = \{u\in V^{(2)}:\ \mbox{$f_v < 0$ contains $\gamma$}\}$.

For each cell $\gamma$,
we recursively construct a biclique cover for the subgraph induced by $V^{(1)}_\gamma
\cup V^{(2)}_\gamma$, the subgraph induced by $V^{(1)}_\gamma
\cup Y^{(2)}_\gamma$, and the subgraph induced by $V^{(1)}_\gamma
\cup Z^{(2)}_\gamma$.
Note that in second and third recursive calls, $\ell$ decreases by 1 
(since the last inequality $f_v(x_{\xi_\ell},x_{\eta_\ell})\ge 0$
has been resolved for $Y^{(2)}_\gamma$ and $Z^{(2)}_\gamma$).  Thus, the biclique cover size satisfies a recurrence
of the form
\begin{equation}\label{eqn:biclique:recur}
 S_\ell(n_1,n_2)\ \le\ b^{2+o(1)} S_\ell(n_1/b^2, n_2/b) + b^{2+o(1)} S_{\ell-1}(n_1,n_2),
\end{equation}
with trivial base cases $S_0(n_1,n_2)=O(n_1+n_2)$ and $S_\ell(1,n_2)=O(n_2)$.
Starting from $(n_1,n_2)=(n,n)$,
the number of subproblems of size $(n_1,n_2)=(n/b^{2i},n/b^i)$ generated by
the recursion is bounded by
$\min\{ (b^{2+o(1)})^{i+\ell},n\}$, and so
$S_\ell(n,n)=O\left(\sum_i \min\{b^{(2+o(1))i},n\}\cdot n/b^i\right)\le O(n^{3/2+\eps})$ for a sufficiently large constant $b$.
Moreoever, each vertex in $V^{(1)}$ participates in $O(\log^{\ell}n)=\OO(1)$
bicliques.  All the properties stated in the lemma thus follow.
\hfill\qed

\ssubsection{Proof of Lemma~\ref{lem:bicliques:2d}}\label{app:bicliques:2d}

We modify the proof of Lemma~\ref{lem:bicliques:2d:0}.
Assume that each range $R^{(12)}(v)\subseteq \R^d$ is expressed as a  fixed Boolean combination of $\ell_1$ polynomial inequalities like before.
Assume also that each range $R^{(21)}(v)\subseteq \R^d$ is similarly  expressed as a fixed Boolean combination of $\ell_2$ polynomial inequalities.
Let $\ell=\ell_1+\ell_2$.
In addition to the earlier recurrence (\ref{eqn:biclique:recur})
on the biclique cover size,
we obtain the symmetric recurrence
\[ S_{\ell}(n_1,n_2)\ \le\ b^{2+o(1)} S_{\ell}(n_1/b, n_2/b^2) + b^{2+o(1)} S_{\ell-1}(n_1,n_2).
\]
Applying the two in succession yields
\[ S_{\ell}(n,n)\ \le\ b^{4+o(1)} S_{\ell}(n/b^3, n/b^3) + b^{4+o(1)} S_{\ell-1}(n,n).
\]
The number of subproblems of size $(n/b^{3i},n/b^{3i})$ generated by
the recursion is at most
$(b^{4+o(1)})^{i+\ell}$, and so
$S_{\ell}(n,n)=O(\sum_{i:\, b^{3i}\le n} b^{(4+o(1))i} \cdot n/b^{3i})\le O(n^{4/3+\eps})$ for a sufficiently large constant $b$.
\hfill\qed

\ssubsection{Proof of Lemma~\ref{lem:bicliques:segs}}\label{app:bicliques:segs}

The proof is based on segment trees~\cite{ChazelleEGS94}.

For a line segment $s$ and a vertical slab $\sigma$,
we say that $s$ is \emph{short} in $\sigma$ if $s$ has an endpoint
strictly inside $\sigma$; we say that $s$ is \emph{long} in $\sigma$ if
$s$ has an endpoint on the left side of $\sigma$ and another
endpoint on the right side of~$\sigma$.

Suppose that we are given a set $S$ of line segments, all short in
a given vertical slab $\sigma$.  (Initially, $\sigma$ is the entire
plane.)  Let $n$ be the number of endpoints strictly inside $\sigma$.  Clearly,
$|S|\le n$.  We want to construct a biclique cover for the intersection
graph of $S$.

Divide $\sigma$ into two subslabs $\sigma_1$ and $\sigma_2$ each containing
$n/2$ endpoints.
Fix $i\in \{1,2\}$.
Clip the segments of $S$ to $\sigma_i$.
Let $S_i$ be the resulting short segments in $\sigma_i$,
and $L_i$ be the resulting long segments in $\sigma_i$.
We recursively construct a biclique cover for $S_i$.
It remains to take cover intersections involving $L_i$.
The intersection graph of $L_i$ is a permutation graph,
and there are known straightforward $\OO(n)$-time algorithms to color
a permutation graph,
yielding a proper $k$-coloring if it does not contain $K_k$, 
or outputting a $K_k$ otherwise.
If a $K_k$ is returned, we can stop.
Otherwise, take each color class $L_i^{(j)}\ (j\in [k])$,
which is a disjoint set of long segments.
Sort them from bottom to top, 
and assign such segment $s\in L_i^{(j)}$ its rank $p(s)$
in the sorted list.
Map each segment $s'\in S_i\cup L_i$ to an interval $I(s')$
consisting of the ranks of all segments in $L_i^{(j)}$ that intersect $s'$.
This is indeed an interval, i.e., a contiguous subsequence of integers, and its two endpoints can be found by binary search.
We can then compute a biclique cover for $\{(s,s')\in L_i^{(j)}\times (S_i\cup L_i): p(s)\in I(s')\}$ of size $\OO(n)$, e.g., by applying Lemma~\ref{lem:bicliques:boxes} in one dimension.

The total biclique cover size satisfies the recurrence
\[ S(n) \le 2 S(n/2) + \OO(n),
\]
which solves to $\OO(n)$.
Moreoever, each segment
participates in $O(\log n)$ nodes of the recursion, and thus participates
in $\OO(1)$ bicliques.
\hfill\qed

\bigskip
Alternatively, one could apply a result from combinatorial geometry by Fox and Pach~\cite{FoxP12}, stating
that line-segment intersection graphs avoiding $K_k$ can be colored
with polylogarithmically many colors (though an algorithmic
version of this result would be needed).  Then for each pair
of colors, one could directly invoke the bichromatic segment intersection algorithm by
Chazelle et al.~\cite{ChazelleEGS94} (of which our algorithm above is a modification) 
to obtain a biclique cover of $\OO(n)$ size.

\IGNORE{

\ssubsection{Proof of Lemma~\ref{lem:segs:I2}}\label{app:segs:I2}

We seek a red segment $s_1^*$ and a blue segment $s_2^*$ that do not intersect.  There are 4 possibilities:
(i)~$s_1^*$ is above the line through $s_2^*$,
(ii)~$s_1^*$ is below the line through $s_2^*$,
(iii)~$s_2^*$ is above the line through $s_1^*$, or
(iv)~$s_2^*$ is above the line through $s_1^*$.
It suffices to consider case~(iv), since the other cases can be handled similarly.

Let $p_1^*$ and $q_1^*$ be the left and right endpoints of $s_1^*$.
Case~(i) can be further broken into two subcases:
(a)~$p_1^*$ is above the line through $s_2^*$ and the slope of $s_1^*$ is greater than the slope of $s_2^*$, or
(b)~$q_1^*$ is above the line through $s_2^*$ and the slope of $s_1^*$ is less than the slope of $s_2^*$.
It suffices to consider case~(a).

\newcommand{\HH}{{\cal H}}
To this end, we maintain the upper hull $\HH_\mu$ of the set of 
the left endpoints $p_1$ of all red segments $s_1$ with slope greater than $\mu$, as $\mu$ decreases from $\infty$ to $-\infty$.
The hull undergoes $O(n)$ insertions of points, and can be
maintained in $O(\log n)$ time per insertion by a known incremental convex hull algorithm~\cite{Preparata79}.  For each blue segment $s_2$ with slope $m$, we just check whether there exists a vertex $p_1$ of $\HH_m$ that is above the line through $s_2$, in $O(\log n)$ time by binary search.  The total running time is $O(n\log n)$.
\hfill\qed

}

\ssubsection{Proof of Lemma~\ref{lem:sparse:C3}}\label{app:sparse:C3}

We modify the $O(m^{2\omega/(\omega+1)})$-time algorithm of Alon, Yuster, and Zwick~\cite{AlonYZ97} or finding 3-cycles.

For a vertex $v\in V^{(1)}\cup V^{(3)}$, call $v$ \emph{low} if 
it has at most $\D$ neighbors in $V^{(2)}$, and \emph{high} otherwise.
For a vertex $v\in V^{(2)}$, call $v$ \emph{low} if 
it has at most $\D m'/m$ neighbors in $V^{(3)}$, and \emph{high} otherwise.
We seek a 3-cycle $v_1^*v_2^*v_3^*$ with $v_1^*\in V^{(1)}$, $v_2^*\in V^{(2)}$, and $v_3^*\in V^{(3)}$.  
We consider two cases:

\begin{itemize}
\item {\sc Case 1:} $v_1^*$ or $v_3^*$ is low.  W.l.o.g., say it is $v_1^*$. We go through each of the $O(m')$ edges $v_1v_3$ in $V^{(1)}\times V^{(3)}$, and if $v_1$ is low, examine each of its $O(\D)$ neighbors $v_2\in V^{(2)}$ and check whether $v_1v_2v_3$ is a 3-cycle.  The running time 
is $O(m'\D)$.
\item {\sc Case 2:} $v_2^*$ is low.  We go through each of the $O(m)$ edges $v_1v_2$ in $V^{(1)}\times V^{(2)}$, and if $v_2$ is low, examine each of its $O(\D m'/m)$ neighbors $v_3\in V^{(3)}$ and check whether $v_1v_2v_3$ is a 3-cycle.  The running time 
is $O(m\cdot \D m'/m)=O(m'\D)$.
\item {\sc Case 3:} $v_1^*, v_2^*, v_3^*$ are all high.
There are at most $O(m/\D)$ high vertices in $V^{(1)}\cup V^{(3)}$,
and at most $O(m/(\D m'/m))=O(m^2/(\D m'))$ high vertices in $V^{(2)}$.
We can detect a 3-cycle among the high vertices by rectangular matrix multiplication in $O(M(m/\D,m^2/(\D m'),m/\D))$ time.
\hfill\qed
\end{itemize}


\ssubsection{Proof of Lemma~\ref{lem:sparse:C4}}\label{app:sparse:C4}

We modify the $O(m^{3/2})$-time algorithm of Alon, Yuster, and Zwick~\cite{AlonYZ97} for finding 4-cycles.

W.l.o.g., say $m'\le m$.
For a vertex $v$ in $V^{(2)}$ (resp.\ $V^{(4)}$), call $v$ \emph{low} if 
it has at most $\D$ neighbors in $V^{(3)}$ (resp.\ $V^{(1)}$), and \emph{high} otherwise.
We seek a 4-cycle $v_1^*v_2^*v_3^*v_4^*$ with $v_1^*\in V^{(1)}$, \ldots, $v_4^*\in V^{(4)}$.  

\begin{itemize}
\item {\sc Case 1:} $v_2^*$ and $v_4^*$ are both low. 
We go through each of the $O(m)$ edges $v_1v_2$ in $V^{(1)}\times V^{(2)}$,
and if $v_2$ is low, examine each of its $\D$ neighbors $v_3\in V^{(3)}$
and mark the pair $(v_1,v_3)$ red.
Similarly, we go through each of the $O(m)$ edges $v_3v_4$ in $V^{(3)}\times V^{(4)}$,
and if $v_4$ is low, examine each of its $d$ neighbors $v_1\in V^{(1)}$
and mark the pair $(v_1,v_3)$ blue.
If a pair is marked both red and blue, we have found a 4-cycle $v_1v_2v_3v_4$.
The running time 
is $O(m\D)$.
\item {\sc Case 2:} $v_2^*$ or $v_4^*$ is high.  W.l.o.g., say it is $v_2^*$.  There are at most $O(m'/\D)$ high vertices in $V^{(2)}$.
From each such vertex $v$, we run breadth-first search to find a 4-cycle through $v$.  The running time is $O(m'/\D \cdot (m+m')) = O(mm'/\D)$.
\end{itemize}

Choosing $\D=\sqrt{m'}$ yields the result.
\hfill\qed

\bigskip
Note that there might be room for improvement in Lemma~\ref{lem:sparse:C4},
by modifying Yuster and Zwick's $O(m^{1.48})$ algorithm for $C_4$ in sparse
graphs~\cite{YusterZ04} instead of the $O(m^{3/2})$ algorithm.
This would require more effort, and the resulting improvement to Theorem~\ref{thm:boxes:K4} would likely be very small.

{\small
\bibliographystyle{plainurl}
\bibliography{geom_cycle}
}

\end{document}